\documentclass[final,5p,times,authoryear]{elsarticle}
\usepackage{lineno,hyperref}
\usepackage{stfloats}
\usepackage{lipsum}
\usepackage{color}
\usepackage{appendix}
\usepackage{listings}
\usepackage{xcolor}
\usepackage[export]{adjustbox}

\modulolinenumbers[5]

\journal{Journal of Astronomy \& Computing}

\usepackage{numcompress}\biboptions{authoryear}

\newcommand{\HI}{H{\textsc i}}
\newcommand{\VR}{\textit{VR}}
\newcommand{\IDAVIE}{\textit{iDaVIE}}

\newcommand{\mnras}{\textit{MNRAS}}
\newcommand{\apj}{\textit{ApJ}}
\newcommand{\apjs}{\textit{ApJS}}
\newcommand{\apjl}{\textit{ApJL}}

\newcommand{\aj}{\textit{AJ}}

\newcommand{\pasa}{\textit{PASA}}
\newcommand{\aap}{\textit{A \& A}}

\colorlet{punct}{red!60!black}
\definecolor{background}{HTML}{EEEEEE}
\definecolor{delim}{RGB}{20,105,176}
\colorlet{numb}{magenta!60!black}

\lstdefinelanguage{json}{
    basicstyle=\normalfont\ttfamily,
    numbers=left,
    numberstyle=\scriptsize,
    stepnumber=1,
    numbersep=8pt,
    showstringspaces=false,
    breaklines=true,
    frame=lines,
    backgroundcolor=\color{background},
    literate=
     *{0}{{{\color{numb}0}}}{1}
      {1}{{{\color{numb}1}}}{1}
      {2}{{{\color{numb}2}}}{1}
      {3}{{{\color{numb}3}}}{1}
      {4}{{{\color{numb}4}}}{1}
      {5}{{{\color{numb}5}}}{1}
      {6}{{{\color{numb}6}}}{1}
      {7}{{{\color{numb}7}}}{1}
      {8}{{{\color{numb}8}}}{1}
      {9}{{{\color{numb}9}}}{1}
      {:}{{{\color{punct}{:}}}}{1}
      {,}{{{\color{punct}{,}}}}{1}
      {\{}{{{\color{delim}{\{}}}}{1}
      {\}}{{{\color{delim}{\}}}}}{1}
      {[}{{{\color{delim}{[}}}}{1}
      {]}{{{\color{delim}{]}}}}{1},
}

\begin{document}

\begin{frontmatter}

\title {Exploring and Interrogating Astrophysical Data in Virtual Reality}

%

\author[1,2,3]{T.H.~Jarrett\corref{cor1}}
\ead{tjarrett007@gmail.com}

\author[2]{A.~Comrie}

\author[1,2,4]{L.~Marchetti}

\author[1,2]{A. Sivitilli}

\author[1,2]{S. Macfarlane}

\author[4]{F. Vitello}
\author[5]{U. Becciani}

\author[2,6]{A.~R.~Taylor}

\author[7]{\,\,\,\,\,\,\,\,\,\,\,\,\,\,\,\,\,\,\,\,\,\,\,\,\,\,\,\,\,\,\,\,\,\,\,\,\,J.M. van der Hulst}

\author[8]{P. Serra}

\author[9]{Neal Katz}

\author[10]{M.E. Cluver}

\cortext[cor1]{Corresponding author}

\address[1]{Department of Astronomy, University of Cape Town, Private Bag X3, 7701, Rondebosch, Cape Town, South Africa}
\address[2]{Inter-University Institute for Data Intensive Astronomy (IDIA), University of Cape Town, Rondebosch,  Cape Town, 7701, South Africa}
\address[3]{Western Sydney University, Locked Bag 1797, Penrith South DC, NSW 1797, Australia}
\address[4]{INAF - Institute for Radio Astronomy, Via Gobetti 101, 40129, Bologna, Italy}
\address[5]{INAF - Catania Astrophysical Observatory, Via Santa Sofia 78, 95123, Catania, Italy}
\address[6]{Department of Physics and Astronomy, University of the Western Cape,
Private Bag X17, 7535 Bellville, Cape Town, South Africa}
\address[7]{Kapteyn Astronomical Institute, University of Groningen, Landleven 12, 9747AD Groningen, The Netherlands}
\address[8]{INAF - Osservatorio Astronomico di Cagliari, Via della Scienza 5, 09047 Selargius, CA, Italy}
\address[9]{Astronomy Department, University of Massachusetts, Amherst, MA 01003, USA}
\address[10]{Centre for Astrophysics and Supercomputing, Swinburne University of Technology, John Street, Hawthorn 3122, Victoria, Australia}




\begin{abstract}

Scientists across all disciplines increasingly rely on machine learning algorithms to analyse and sort datasets of ever increasing volume and complexity. Although trends and outliers are easily extracted, careful and close inspection will still be necessary to explore and disentangle detailed behaviour, as well as identify systematics and false positives. We must therefore incorporate new technologies to facilitate scientific analysis and exploration. Astrophysical data is inherently multi-parameter, with the spatial-kinematic dimensions at the core of observations and simulations.  The arrival of mainstream virtual-reality (VR) headsets and increased GPU power, as well as the availability of versatile development tools for video games, has enabled scientists to deploy such technology to effectively interrogate and interact with complex data. In this paper we present development and results from custom-built interactive VR tools, called the \IDAVIE\ suite, that are informed and driven by research on galaxy evolution, cosmic large-scale structure, galaxy-galaxy interactions, and gas/kinematics of nearby galaxies in survey and targeted observations. In the new era of Big Data ushered in by major facilities such as the SKA and LSST that render past analysis and refinement methods highly constrained, we believe that a paradigm shift to new software, technology and methods that exploit the power of visual perception, will play an increasingly important role in bridging the gap between statistical metrics and new discovery. We have released a beta version of the \IDAVIE\ software system that is free and open to the community.

\end{abstract}


\end{frontmatter}

\section{Introduction}

Data visualisation plays an important role in the analysis and dissemination of scientific data, which is increasingly becoming more complex and large in volume. Our primary tools of the trade, computers (hardware and software, thereof), may in many ways decrease the burden of data `reduction' for the scientist as they carry out the calculations by which we transform raw data into refined products and meta information. 

And yet the most crucial step in the scientific method, analysis (and subsequent interpretation), cannot solely rely on computers and automated algorithms, e.g. machine-learning tools -- which are only as good as the programming and input (``truth") knowledge that drive them -- to make the critical breakthroughs that are hidden or obscured by the multi-dimensional dependencies.  These discoveries are typically few and far between.  It is through visualisation and critical analysis that revelations of underlying and nuanced knowledge and understanding are made; and in the context of machine learning and neural-network algorithms, visualisation is often the most reliable way to prune training sets, and can be employed to understand the complex systems that go into mapping these networks \cite[cf.][]{Gal03}.

It is this step, analysis and validation of the data, for which we bring to bear our most human of tools -- intuition and creativity -- that is largely informed by graphical visualisation, exploiting the human visual perception of reality \citep[cf.][]{HealeyEnns,McC1987}.
Graphics may range from simple scatter plots (from 2 to N dimensions), to multi-color histograms, charts, imaging, 3D (volumes) and dynamic rendering (i.e., videos).  All are designed to be optimally effective with the way humans visually process information. Usually this is done by looking at a flat, 2D computer screen, which in fact limits the evolved visual skills of humans; notably, the ability to use all of our vision to assimilate and perceive within our natural 3-dimensional space and, of course the dynamic fourth, time.  This is especially a challenge with 3D and multi-dimensional data sets projected onto a flat screen.
 Fortunately, with modern technology our graphics can now move beyond this limitation.

\begin{figure}

\includegraphics[width=0.48\textwidth]{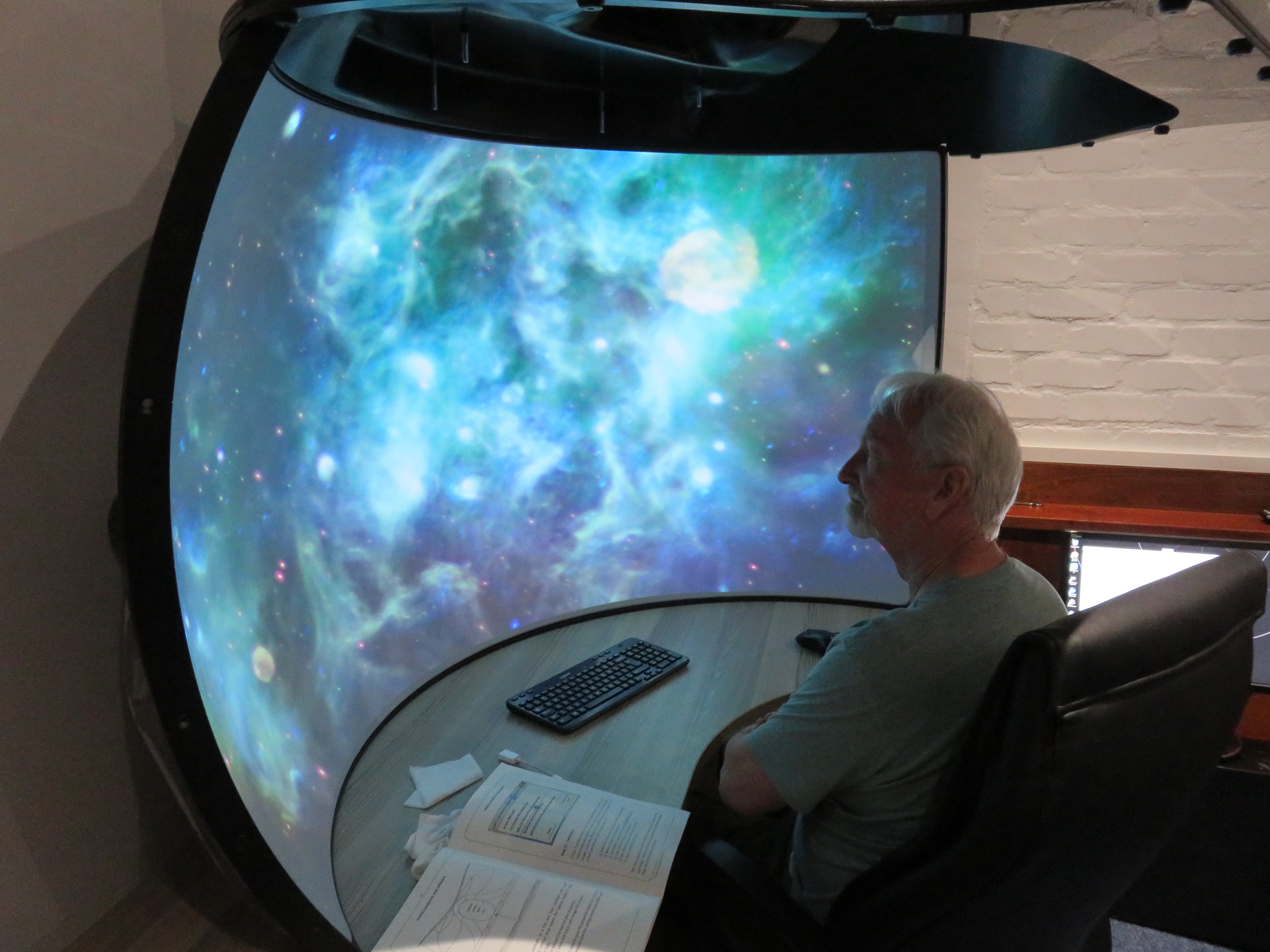}
\centering
\vspace{-10pt}
\caption{\footnotesize{
Immersion technology has changed the way we view our data.  
A very effective device for viewing 3D astrophyscial data is the Elumenti Cobra, shown in action here in the IDIA Visualisation Laboratory (IVL). 
}}\label{fig:cobra}
\vspace{-10pt}
\end{figure}

With the advent of graphical processing units (GPUs), immersion technology has enabled a far more powerful and natural way to visualise complex data, whether it comes from an instrument experiment or is derived from computer models and simulations.  Immersion comes in the form of curved screens and monitors (e.g., see Fig~\ref{fig:cobra}), and the latest digital-projection planetariums are increasingly being used for scientific visualisation.
Early efforts to visualise and conduct research with astrophysical data showed the promise of immersion and virtual reality technologies, which included head-sets,  walls and cylinders (e.g., CAVE), 
and  full-dome (digital 360 degree planetarium) facilities 
\citep[cf.][]{Hassan2011,
Djor2013, FEI2016, FB2018, LMarch2018}.
More recent efforts have added utility and sophistication to the such methods, moving closer to practical tools that can be deployed across the full astronomy research spectrum, that also includes sophisticated simulations and numerical modeling 
\cite[see][]{Davelaar2018, Dykes2018, BV2019, Marchetti20, Dykes21}

One of the most promising technologies is so-called ``virtual reality''  (which we henceforth refer to as \VR), which places the human in an artificial 3D environment along with their data (or their ``game", as \VR\ was originally created for gaming); in other words, a 3D ``monitor" which envelops the user.  The scientist is no longer limited by 2D projections, but is able to move within, interact, and manipulate the data using natural motions and gestures.  In the last few years there has been much research into how \VR\ and augmented-reality (AR) is changing the way we interact with our data (analysis) and with other humans (socialisation, entertainment, communication, etc), investigating such aspects as visual perception, haptic feedback, auditory perception, pain management,  task and learning performance, psychophysics, bio-medical imaging and surgery, and just about any kind of human interactive experience
\citep[to name a few works, see][]{Jin2012, HCC2016, Hoff2014,Shatt2018,SP2018,VQ2021}.



In this study, we focus on \VR\ and the application to astrophysical data.  We start with a more descriptive definition of \VR, as given by \citet{Rubio2017},
quoting: 
\emph{A medium composed of interactive computer simulations that sense the
participant’s position and actions and replace or augment feedback to one or more senses, giving the
feeling of being mentally immersed or present in the simulation (a virtual world)}.
Any data that is multi-dimensional -- which is to say, most kinds of scientific research data -- lend themselves well to the \VR\ environment because of this immersion feature.  Astrophysics is a perfect example; astronomers study in detail the internal structure of stars, nebulae, galaxies and the Universe itself (e.g., cosmic large scale structure, or the Cosmic Web).  All of these are 3D and dynamic (i.e., time dimension), and traditionally analysed in sliced or projected 2-dimensions, greatly simplifying but also limiting the view of the complex systems we are trying to disentangle and understand.  Immersive technology opens up so many possibilities and new avenues to explore our rich -- and exponentially expanding -- multidimensional data sets.



\begin{figure}[h!]
\includegraphics[width=0.48\textwidth]{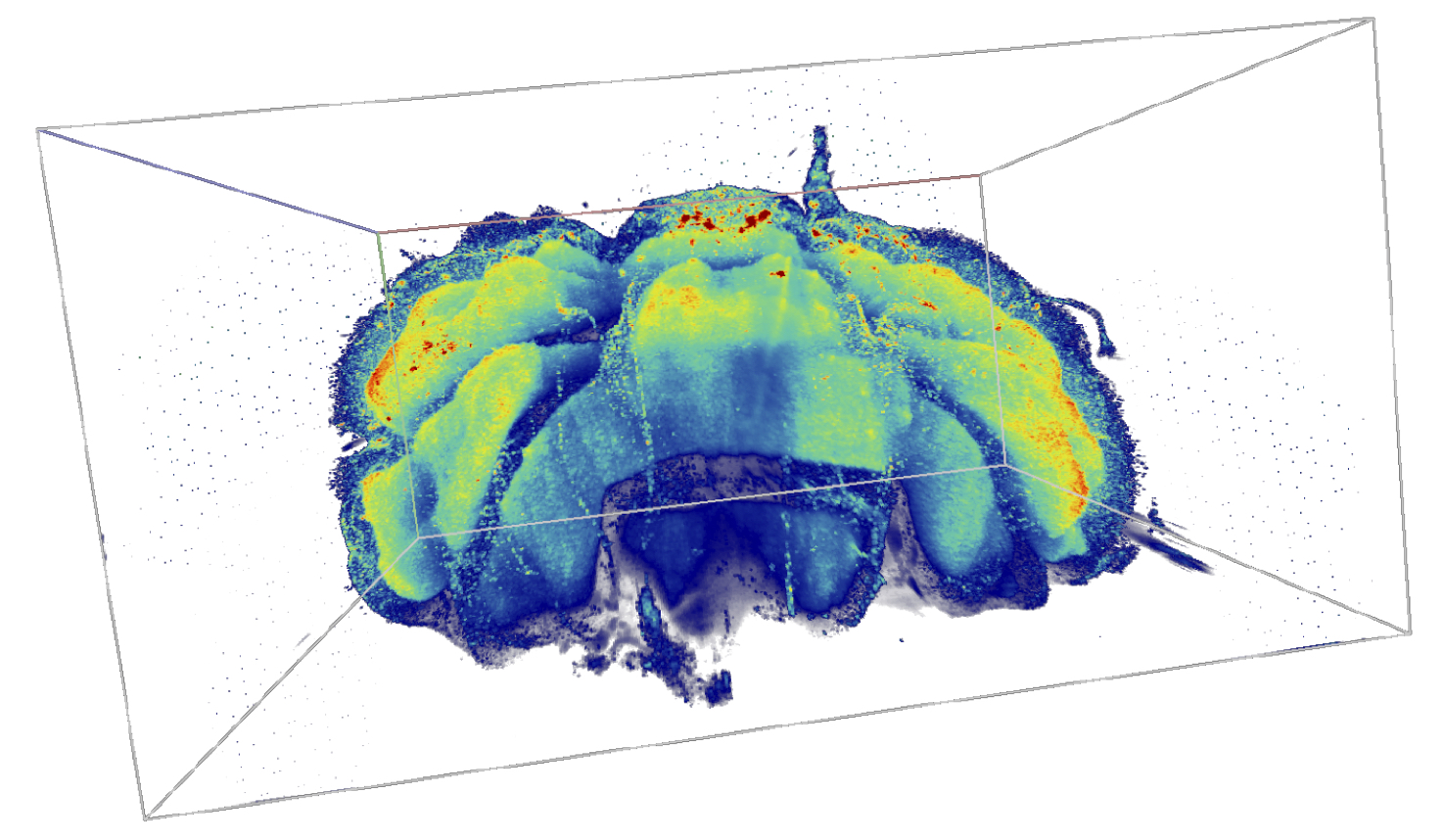}
\centering
\vspace{-20pt}
\caption{\footnotesize{
Mouse cerebellum rendered in our \VR\ system.  The color transform highlights 
sensory system tissue which is involved in
the coordination of voluntary movements such as posture and balance, resulting in smooth and
balanced muscular activity. Processed using CLARITY, a method that renders tissue transparent by
washing out light-scattering lipids and allowing for 3D fluorescent imaging of large tissues. A Carl Zeiss
LSM780 confocal microscope was used for tissue imaging. Data from B. Loos \& A. Du Toit (Stellenbosch Univ.).
}}\label{fig:mouse}
\vspace{-5pt}
\end{figure}

The next step in this evolutionary process, as the \VR\ and GPU hardware becomes ever more advanced, is to develop interactive software tools. 
As summarized by \cite{BV2019}, to quote: 
\emph{
We conclude that the future of \VR\ for scientific purposes in astrophysics most likely resides in the development of a robust, generic application dedicated to the exploration and visualization of 3D observational datasets, akin to a ds9-VR}.
Indeed, the development in our lab is based on this primary objective, to create \VR\ tools that allow and enable interaction with scientific data.  

A confluence of hardware improvements and data complexity imperatives have propelled \VR\ development.  Another equally important arrival are 
cadence-driven mega surveys (e.g., LSST) and the long-wavelength 
interferometers, notably in the mm-wave (e.g., ALMA) and the radio-wave 
(Square Kilometer Array (SKA) international initiative and its pathfinders.
These massive interferometers, including Australia's ASKAP and South Africa's MeerKAT,
are producing spectral (velocity) cubes with enormously larger volumes.  Intrinsically $>$2D in nature, these data products require a re-evaluation as to how we make sense of the copious information that they provide.  With this as our research imperative and development driver, we created the 
 IDIA Visualisation Lab (IVL)\footnote{IVL: https://vislab.idia.ac.za/} at the University
of Cape Town, established as a joint project between UCT Astronomy and the
Inter-University Institute for Data Intensive Astronomy
(IDIA).

\begin{figure}[h!]
\includegraphics[width=0.48\textwidth]{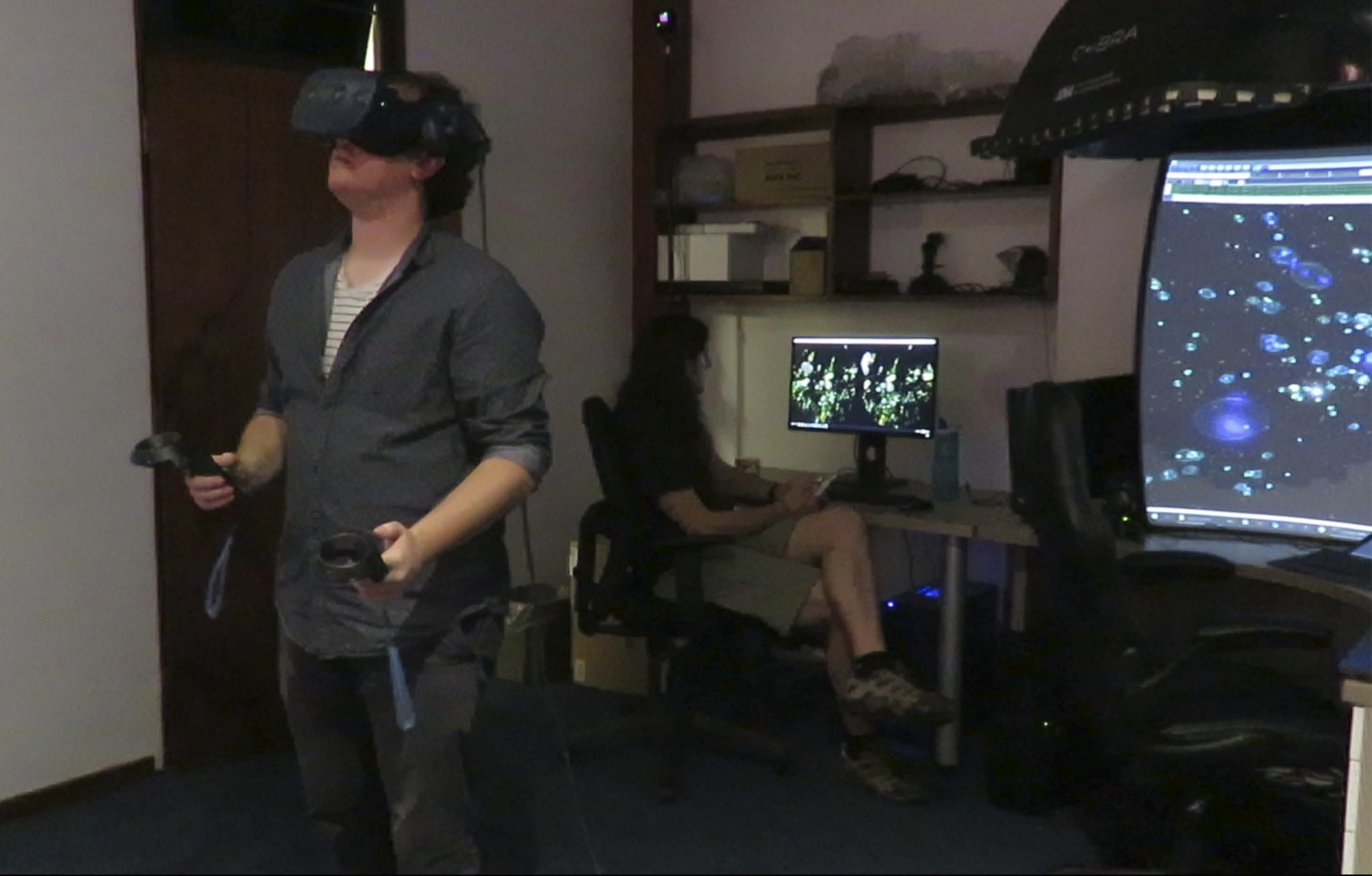}
\centering
\vspace{-15pt}
\caption{\footnotesize{
Using the \VR\ system described in this work, research on the large scale structure of galaxies from the 2MASSS Redshift Survey (2MRS) is carried out by masters student, Trystan Lambert (see \S\,4).
}}\label{fig:vislab}
\vspace{-5pt}
\end{figure}
Since our founding in 2017, 
we have developed visualisation tools in our lab, deploying a number of immersion devices across a diversity of data sets and platforms (see  Fig~\ref{fig:cobra}), as well as across disciplines (see for example the 3D rendering of a mouse cerebellum using our system;  Fig~\ref{fig:mouse}).  
The majority of our work thus far has focused on \VR\ exploration of astrophysical data, 
both in the form of source catalogues -- featuring for example cosmic large scale structure of galaxies, see Fig~\ref{fig:vislab} --  and 3D spectral imaging (cubes) -- for example in the form of atomic (neutral) hydrogen emission in galaxies, showcased in Section 4.  Our development is
largely driven by the major radio interferometry science -- through the SKA Initiative -- that is rapidly accelerating in South Africa, Australia, India, and Europe, and is a major focus of this presented work.



In this paper, 
we start  (\S\,2) with the science cases that have informed and driven the IVL development that is the subject of this document.   With this context in place, we then (\S\,3) present the details of our \VR\ software suite: 
\IDAVIE\ (Immersive Data Visualization Interactive Explorer), detailing 
the standard techniques that we deploy (e.g., ray marching), and the novel methods we  have developed to optimize working with data in \VR.  In (\S\,4), we present 
 the science results from astrophysical projects that use our system for rendering particles and volume data sets for 3D exploration and interaction.  We conclude the section with our extensive development working with volumetric data, notably spectral ``cubes" derived from radio interferometry.
 Finally, \S\,5 discusses the lessons learned and continuing challenges faced by \VR\ for research purposes, and considers the path forward as we see development in the next five years.
 We hope to show that the immersive perspective provided by our \IDAVIE\ suit enhances our visual perception and provides a new and powerful avenue for data discovery.   In the examples presented in this paper,  the color libraries available are limited to those in Matplotlib, although in some cases more optimal color transforms may be advised.


\section{Data and Science Drivers}
The UCT-IDIA Visualisation Laboratory (IVL) was created to respond to the challenges of the Big Data Era, and notably from the international LSST and Square Kilometer Array (SKA) initiative and its pathfinders.  Large area surveys producing GB-to-TB sized imaging polarimetry maps, and 3D spectral volumes of emission lines (e.g., \HI, neutral hydrogen at 21\,cm) present a number of data challenges, including storage, pipeline reductions, mosaicking, and most relevant to this study, visualisation and comprehension of the vast and often complex phenomena we observe in the universe.  

\setlength{\tabcolsep}{12pt}
\renewcommand{\arraystretch}{1.25}

\begin{table*}[t]
\centering
\begin{tabular}{ |p{4.1cm}||p{4.3cm}|p{0.75cm}|p{5.4cm}|  }
 \hline
 \multicolumn{4}{|c|}{Table 1:  \VR\ Data and Modes Presented in this Study} \\
 \hline
 Project & data type & mode & comment\\
 \hline
 2MRS Galaxies and Groups (large scale structure)  &  catalogues (VOT/xml, IPAC)   & p &   relatively small number of particles\\
 Galaxy-Galaxy interaction & FITS tables & pt & includes time domain(t)\\
 Cosmological simulation & FITS tables & p & dense numbers of particles\\
 Andromeda Galaxy & FITS cube \& JPEG & v &  volumetric spectral cube and plane\\
 Fornax Galaxy Cluster  & intensity and mask cubes, SoFiA text file & v, d & volumetric spectral cube and ancillary information from SoFiA\\
 \hline
\end{tabular}\label{table1}
\vspace{-10pt}
\end{table*}

Scientific research with virtual reality was identified early on as a priority for hardware and development investment because of the multi-dimensional nature of the data and the science drivers that researchers associated with the lab, or in collaboration with the IVL, were most interested in.   Ranging from 3D catalogues, cosmological simulations, and spectral imaging, 
a summary of the science projects, associated data sets, and the \VR\ mode with which they are deployed, is given in
Table 1.  The modes are ``particle" (``p") and ``volume" (``v") are detailed in \S\,3 and the results discussed in the sections to follow (\S\,4), and finally, a preview of the VR to Dome (``d") mode (\S\,5) we will deploy in the future.

\begin{figure}[h!]
\includegraphics[width=0.48\textwidth]{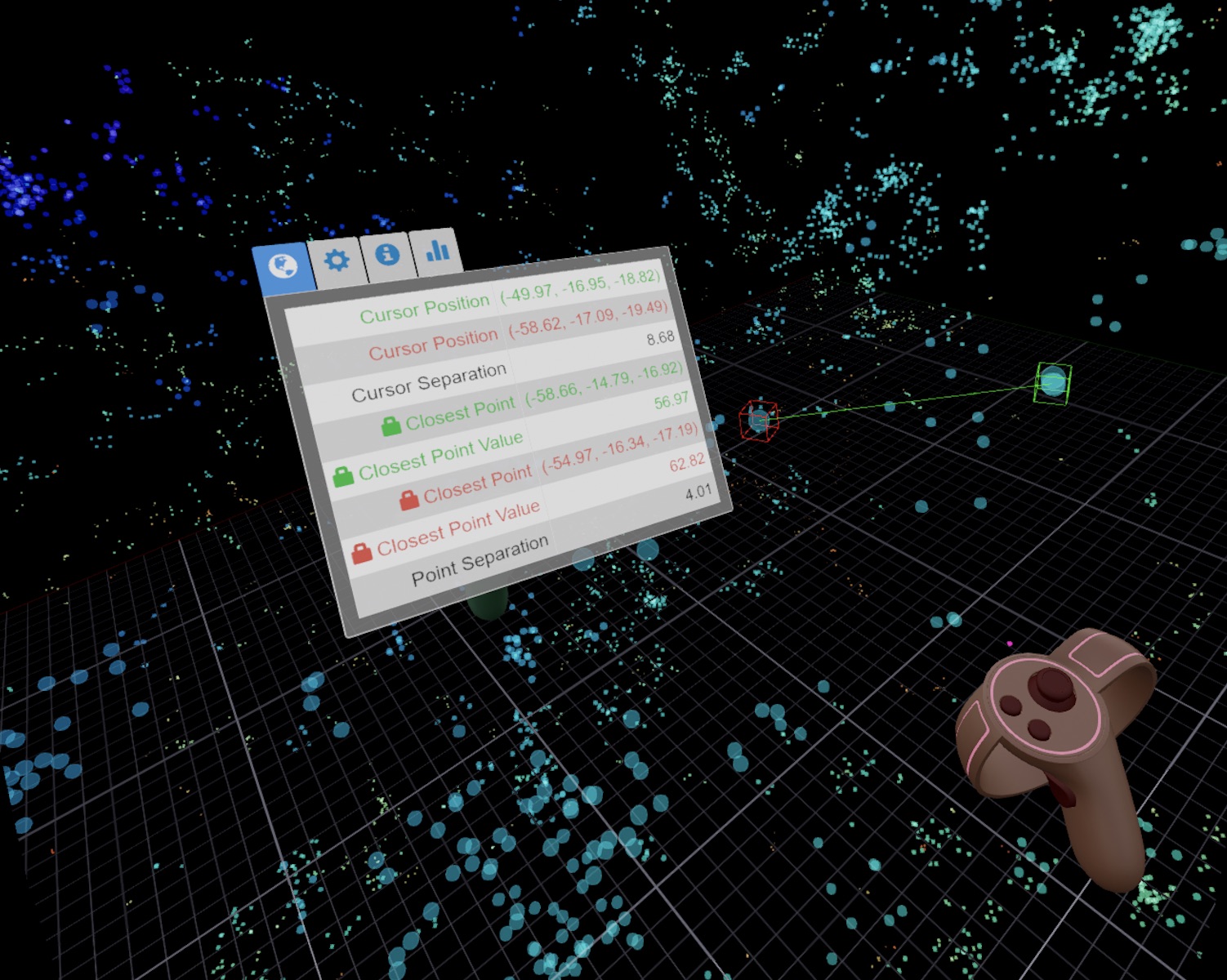}
\centering
\vspace{-15pt}
\caption{\footnotesize{
The 2MRS redshift catalogue 
\VR\ rendered in the \IDAVIE--p mode.  The particles (galaxies) are given a color (rainbow transform) based on its redshift distance.  Here the physical distance (in Mpc) between two galaxies is computed on the fly, demonstrating the user-interaction capabilites in 3D space.  
}}\label{fig:mode-p}
\vspace{-5pt}
\end{figure}

\subsection{Particle Rendering}
Source catalogues are one of the most common data products in the field of astrophysics.  Typically consisting of source coordinates, measured and derived physical attributes.  These are called ``particle" (\IDAVIE--p) data types because they are single objects or entries in a table.
In addition to empirical observations, mock catalogues are often derived 
from theoretrical and semi-empirical model simulations, and are used as a toolset  to inform our understanding of the data.     
These multi-parameter data and simulation sets are usually visualized with scatter plots, histograms and images in 2D; 
but increasingly,  3D is being used as the data quality and the visualisation methods have improved.  A prime example are the redshift surveys.

Since the 1980's and the first redshift surveys, 3D galaxy catalogues have driven key studies of the large scale structure of the universe, the so-called    
 `Cosmic Web' of galaxies.
 A number of projects carried out by IVL associates are doing front-line research in this field, both through empirical redshift surveys and through numerical simulations.  Unlike the spectral cubes, which are volumetric data types, working with galaxy space-distributions is using the sparse particle data type in \VR, which requires slightly different approaches. 
 Fig~\ref{fig:mode-p} illustrates how a 3D particle data set is rendered in \VR, enabling intuitive and easy interaction with the data.
 In \S\,4 we present detailed \VR\ exploration with the 2MASS Redshift Survey (2MRS), and with a cosmological simulation of the local universe, showcasing both empirical and theoretical 3D data sets.

Another particle-type project that was explored by the IVL and associates is n-body simulations, specifically galaxy-galaxy gravitational interaction.  Here the added complexity was the 4th dimension of time (type ``pt"), where the research focused on the 
dynamic evolution of interacting galaxies.  This data set presented 
novel \VR\ challenges (memory, latency, efficiency) to overcome. More details and results are presented in \S\,4.

\begin{figure}[h!]
 \vspace{-5pt}
\includegraphics[width=0.49\textwidth]{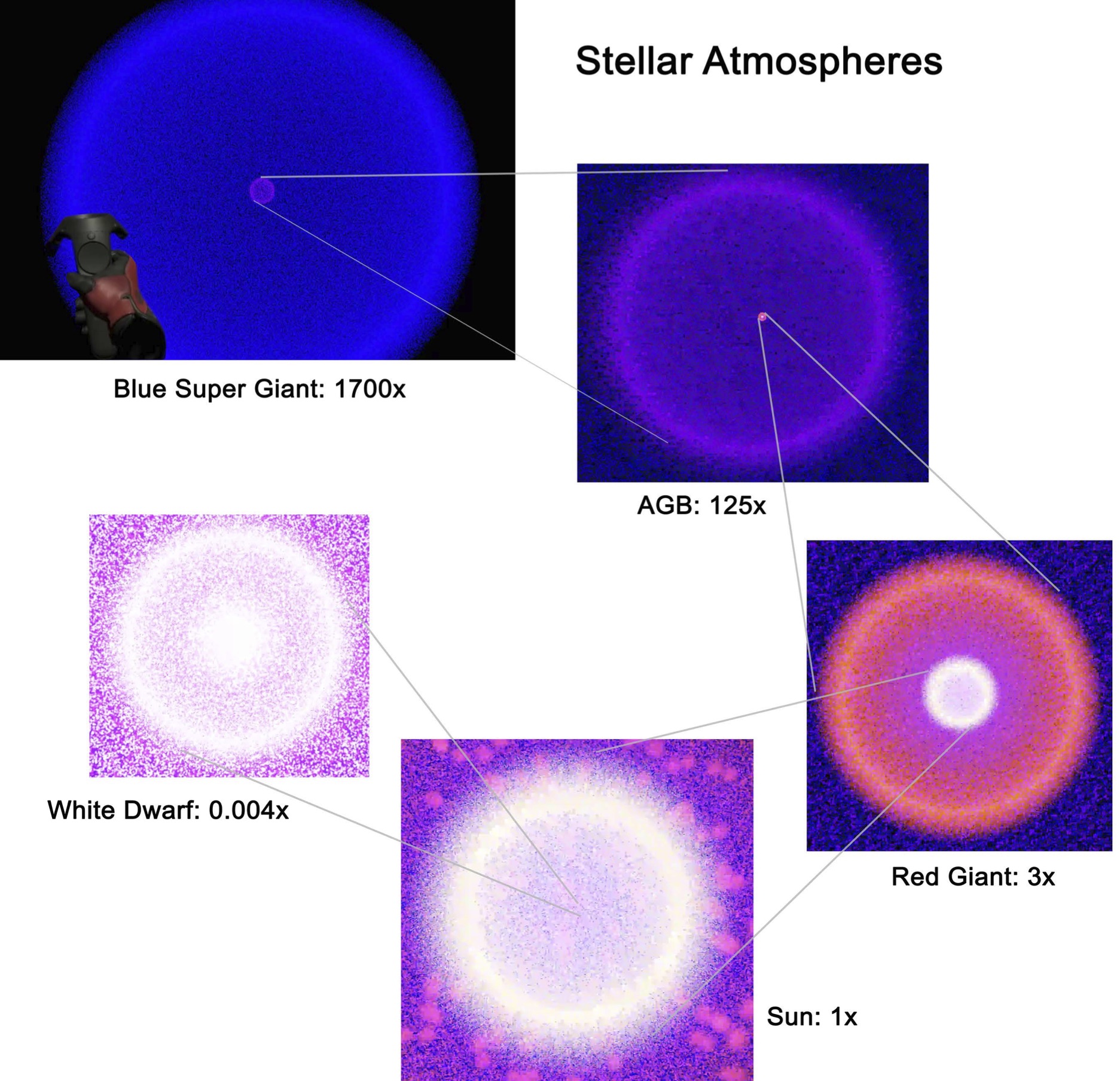}
\centering
\vspace{-20pt}
\caption{\footnotesize{
\VR\ rendering of stellar atmospheres, ranging in size and densities from white dwarf, solar-type, red giant, AGB and blue supergiant modeling. Data (courtesy of M. Joyce) from particle maps generated with the MESA2HYDRO stellar structure-to-particle map interface \citep{Joyce19}; . 
}}\label{fig:atmos}
\vspace{-5pt}
\end{figure}

As a last example of catalogue exploration, and serving as a visual demonstration of physical scales and particle densities, we worked with colleagues \citep{Joyce19} who study the evolution of stars through
stellar atmosphere modeling, from white dwarf to blue supergiants.
The spherical atmospheres, rendered with small particles to convey density, was presented in the \VR\ environment using accurate relative scales (e.g., a red giant star is three times larger than a solar-type star), and colours that are intuitively mapped to the star (e.g, solar spectral types appear yellow).     
 Fig~\ref{fig:atmos} highlights key features of this data set rendering (although severely limited by what can be conveyed in the 2D graphic shown here).

For all these particle types, the input data are stored in multi-column tables, either FITS tables (most efficient) or the VOT/xml standard, unless otherwise noted (e.g., during early development, we used simple ascii text).

\subsection{Volume Rendering}

Many data sets are best rendered as 3D images, or simply volumes. A regular grid in X-Y-Z space is used, with the individual volumetric pixels called ``voxels".  As opposed to sparsely populated (in 3D space) catalogues,
the type of data that is best suited for volumetric rendering is densely packed, filling a large fraction of the data set with information.  A classic example is the spectral image, where the spatial dimensions are the traditional image, and the third orthogonal dimension is spectral.  The spectral information may be frequency, equivalent wavelength or radial velocity (e.g., specific to an emission line).  It is not just astronomy of course, other fields of science and technology utilize spectral imaging, as well as true 3D spatial imaging; e.g., CT and MRI scans in the health industry, tissue probes in biochemistry (see for example, Fig~\ref{fig:mouse}). Hence, it is of special importance to our immersion technology development to have user-interactive functionality in our \VR\ system.   

\begin{figure}[h!]
\vspace{-0pt}
\includegraphics[width=0.49\textwidth]{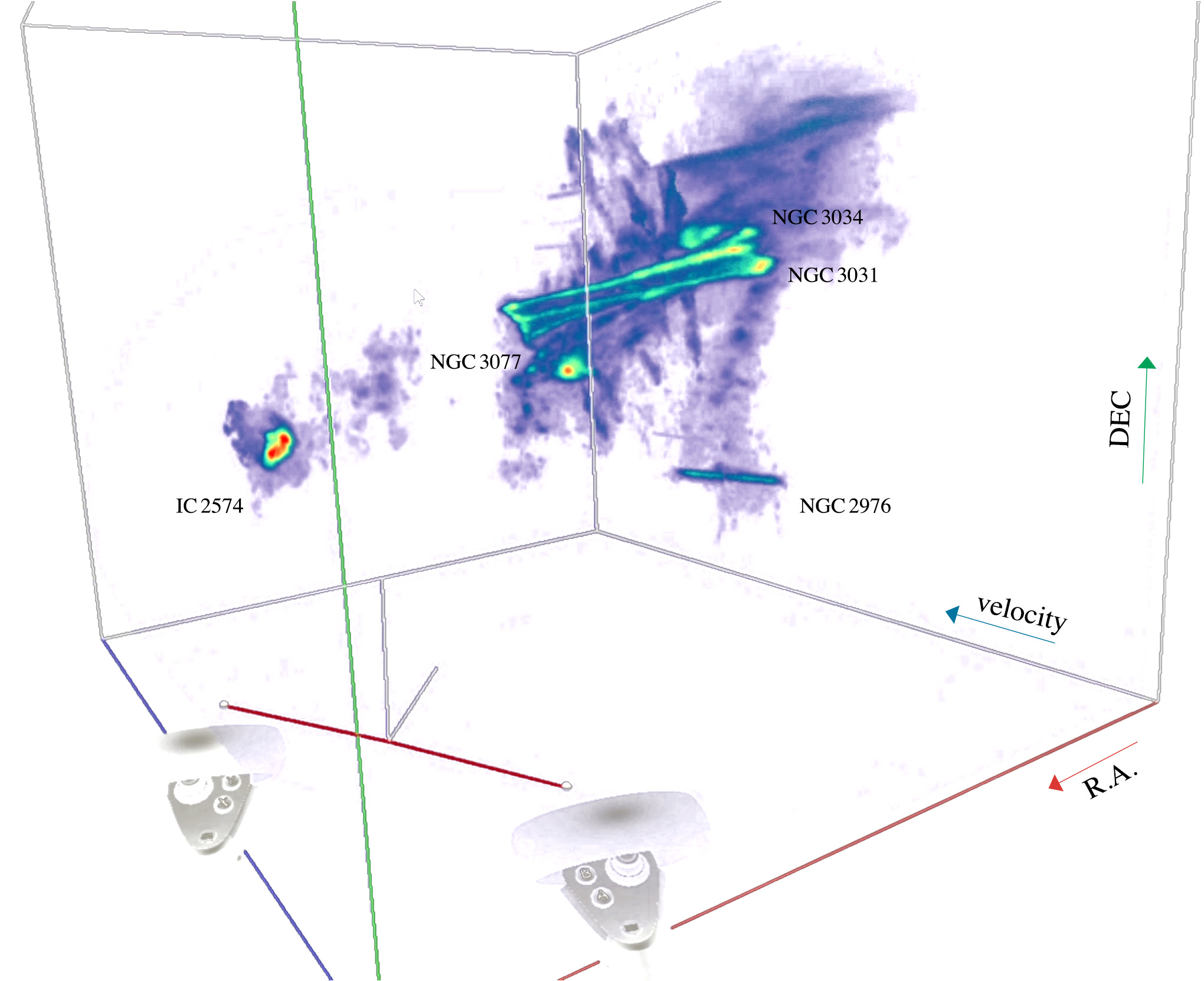}
\centering
\vspace{-15pt}
\caption{\footnotesize{
Atomic hydrogen in the M81 Galaxy Group, viewed in \VR\ with \IDAVIE--v.  The spectral cube shows the spatial (RA-Dec) and velocity distribution of the gas in the group, consisting of galaxies M81 (central), M82 (starburst), NGC\,3077,   NGC\,2976, and IC\,2574, as well as swirling inter-group gas streams from the tidal interaction. Data from \citep{Sorgho19}.
}}\label{fig:m81}
\vspace{-5pt}
\end{figure}
 
A key science driver 
is working directly with 21-cm \HI\ neutral hydrogen.
Elemental hydrogen is the most abundant element in the universe, existing in one of three states:  neutral, ionized and molecular.  The most common is neutral, \HI, which is also the most versatile for study of the cosmic baryon cycle. The enigmatic \HI\ reveals itself when it changes energy state through its orbiting electron spin-flip transition.  At a truly leisurely pace of once per 10 million years, this rare transition is (thankfully) offset by the shear number of hydrogen atoms -- 90\% of all baryonic particles in the universe -- making the 21-cm photons (1.420 Ghz) a prime emission line for radio telescopes. 
With the advent of the first radio interferometers (the WSRT and the VLA in the 1970's, and ATCA in the 80's), and most recently the SKA pathfinders, imaging the \HI\ line in the Milky Way and in galaxies is the most viable way to study the fuel behind all cosmic growth in the universe \citep[see e.g., the seminal work from][]{Allen73,Haschick82}.  
For radio astronomers and the new generation of imaging telescopes, spectral cubes are one of the primary data products. 

We have developed our \VR\ system to work with data products from \HI\ imaging, in particular, but 
our system works with any kind of spectral cube (e.g., IFUs) and any kind of emission line, or even continuum.  Galaxies with rotating disks will exhibit gas that is at lower and higher velocities relative to the central, or systemic velocity, V$_{sys}$.  It is this kinematic information that is contained in the spectral cubes and rendered by our \VR\ system using a ray-marching transfer function method (described in detail, Section 3).  

In the IVL, we began our volumetric development by first looking at spectral cubes of nearby galaxies, including those in the spectactular M\,81 group,  with wide-field data acquired using interferometry from the KAT7 and DRAO synthesis telescopes \citep{Sorgho19}.
 Fig~\ref{fig:m81} illustrates volume rendering of an \HI\ spectral cube using the \IDAVIE--v mode.  Here a wide area of the M81 Group of galaxies is viewed in the spatial plane (X-Y or RA-DEC), and the frequency (or kinematic velocity) in the 3rd (or Z) axis.  This outstanding and complex physical system demonstrates the power and value-add of volume rendering -- the gas kinematics (stretching along the z-axis) is due to both the individual member disk rotations and the multiple-body gravitational interaction between the group members,  creating a stream of gas in the X-Y-Z volume that is nearly inseparable in two dimensions.  
 
 Well-resolved both spatially and kinematically, nearby galaxies put a premium on the contrast between high surface brightness emission (galaxy disks) and fainter diffuse gas emission that is located in the outskirts and in the inter-group medium.  Early \IDAVIE--v development focused on color transforms (of the voxel intensity), adjustments of the minimum and maximum intensity mapping, navigation and volume size optimization. For the M81 Group case, Figure \ref{fig:m81}, balancing the bright emission in the central nuclei and spiral arms with the more diffuse streams between the group members,
\IDAVIE\ defaults to perceptually uniform colourmaps,  such as viridis, but the user can easily switch to alternatives such as jet/rainbow, or its improved variation ``turbo-jet" (as shown in figure), to enhance contrast as needed.
  We explore more detailed examples of volume rendering in \S\,4.


 The spectral cubes are assumed to be in FITS format \citep{wells1979fits}, following the conventions thereof for the headers (importantly, the WCS) and the binary data structure.   Nearly any 3D data, including those from other disciplines, are easily converted to FITS format using conventional tools, but for the purposes of this work presented here, we focus only on the FITS format.
 Detailed results are presented in \S\,4, focusing on \HI\ emission and kinematics of nearby galaxies.

\subsection{Interdisciplinary Application}
We have focused our attention on astrophysical applications of \VR\ exploration (results presented in \S\,4), but it is worthwhile to note that our tools have been developed to be
agnostic to the subject or source of particle and volumetric dataset. 
 It was our intention when we established the IVL that we would work with our science and engineering colleagues,  their data and methods/analysis are, in many ways, closely aligned to those in the astronomy and physics realm. 

Specifically, 
working with our cellular and bio-chemistry colleagues, data in the form of specimen slices and  3D scans was considered and folded into our \VR\ development.  The science drivers included study of neuro-degenerative diseases and cancers.  

The primary challenge of interdisciplinary research is the diversity in data formats.   Astronomers adopted the FITS standard decades ago, which is widely in use, and are slowly moving to the VOT/XML standard for tables.  That makes it easy to control and mitigate the input data development.  Whereas outside of the astronomy community, there are many other standards for data products. In the case of CT, MRI and scanning microscope imaging, the formats may widely vary; e.g., to read in and 3D render the mouse cerebellum scan in \VR\ (see Fig~\ref{fig:mouse}), transformation of the raw 3D TIFF file into the astronomical FITS format (header and binary) was required.  

This cross-disciplinary aspect of \VR\ research is still in a preliminary stages, in this paper we do not further detail interdisciplinary work.  However, in the final section (\S\,5), we discuss the way forward with \VR, notably with streaming technology, which we believe will have wide application across scientific, engineering and health-field disciplines.

 


\section{Development of \IDAVIE}

The aim of \IDAVIE\ (Immersive Data Visualization Interactive Explorer),
is to render datasets in a room-scale 3D space where users can
intuitively view and uniquely interact with their data in ways unrewarded by conventional
flatscreen and 2D solutions. Viewing includes both the conversion of datasets
of typical machine-readable formats to practical representations, along with the ability
for the user to navigate spaces of the virtual setting to see the data from multiple
translational, rotational and scalable viewpoints. Unique interaction with the data in
\VR\ entails modifying viewing parameters, taking measurements, and annotating in the
same space where the data is rendered, effectively allowing the user to remain in the
\VR\ session and perform science directly on the data.

The added complexity of supporting a variety of hardware configurations and \VR\ devices can be prohibitive for small teams. We therefore chose to utilise established game engines, such as Unreal Engine or Unity\footnote{\url{https://www.unrealengine.com} \& \url{https://unity.com/}}, rather than develop \IDAVIE{} from scratch. Both Unreal Engine 4 (UE4) and modern versions of Unity support a number of \VR\ software development packages. We chose to use the SteamVR platform, in order to support a number of popular \VR\ headsets, and eventually extend \IDAVIE{} to support Linux and MacOS. While MacOS support for SteamVR has subsequently been discontinued, Linux support remains a goal, and \IDAVIE{} is designed to use cross-platform libraries to enable this.

Initial work on \IDAVIE{} began with UE4, as we were already familiar with the platform, and the engine's open source C++ codebase allows for easy integration of existing C/C++ scientific packages. However, the complexity of UE4's rendering system made it difficult to quickly develop and integrate GPU shaders into the growing and increasingly complex IDAVIE system. Unity offers an easier route to integrating custom shaders, written in the High Level Shader Language (HLSL). After 6 months of development, we migrated to Unity, allowing us to quickly develop compute and graphical shaders for both the catalogue and volume rendering features. However, the switch to the .NET-based Unity engine came with two cost issues: Firstly, performance-critical code (and code handling datasets larger than 2 GB in size, due to a .NET framework limitation) still needed to be written in C++. Secondly, the majority of scientific packages we utilise were C-based. In order to overcome these issues, we make use of Unity's native code plugin system. A C++ data analysis plugin is used to efficiently perform compute-intensive tasks in parallel using OpenMP, while wrapper plugins were used to interface with scientific libraries such as CFITSIO \citep{Pence99} and AST \citep{Berry16}. 

\subsection{Visualizing Particle Datasets}
\label{rendering:catalogue}
We now describe the \IDAVIE--p mode, relevant to 3D catalogues and
multi-parameter datasets.
Catalogue datasets are sparse in nature, and each row in the catalogue table is represented by a single rendered ``particle'' (rendered as a simple 2D shape, oriented to face the user). Particles can have a number of renderable properties: position, radius, colour, opacity and shape. As many datasets have tens or hundreds of columns, some level of user configuration is required, in order to specify which columns to utilise when rendering, and how to map the values stored in these columns to renderable properties. This user configuration is in the form of a JSON configuration file, which contains information on which column to use for each renderable property, and how to transform the column values to render properties. For example, the configuration file can specify a particular column to use as the input for the particle colour, and choose bounds between a minimum and maximum to scale input values, and a scaling type (linear, logarithmic, square-root or power). The configuration file can also specify uniform properties, if no mapping to a column is required.  See Appendix A for example JSON configuration files that work with the \IDAVIE\ system.

Catalogue datasets are parsed from disk into system memory, where we store all numeric columns in the table. Those columns that are referenced in the configuration file are uploaded to the GPU as 32-bit floating point buffers. A three-stage GPU shader pipeline is then used to transform those data points into rendered pixels:
\begin{enumerate}
\item \textbf{Vertex shader}: The vertex shader stage looks up data from the column buffers, based on the mapping configuration and vertex ID. It then performs the transformation of data values to positions, radius, shape, opacity and colour properties. Colour properties can either be directly determined from the data values, or from a colourmap lookup texture. The resultant vertex information is passed to the next stage of the pipeline.
\item \textbf{Geometry shader}: The geometry shader stage uses the input vertex's position and radius to draw a user-facing quadrilateral of the appropriate size. While this approach (commonly known as ``billboarding'') is not novel (see e.g., \cite{Hassan2011}),
we made some adjustments to it for \VR: instead of assuming the screen is flat, we assume the user's screen is curved, and as such the direction that quadrilaterals must be rotated in order to face the user depends not just on the user's viewing direction, but also the quadrilateral's position in the field of view.
\item \textbf{Fragment shader}: The fragment shader stage fills the quadrilateral with the appropriate texture (according to the particle shape value) and shades it according to the particle colour.
\end{enumerate}

As all the mapping from data values to rendered particles is performed on the GPU, the mapping configuration can be altered in real time for instant feedback. For example, a user can dynamically adjust the particle size without any data reprocessing.

Fully opaque fragments are generally rendered using a \textbf{ Z-buffer} in order to occlude those fragments that are overlapped by fragments that are closer to the camera. Each fragment is tested against the Z-buffer, and only those fragments that have a Z value closer to the camera than the existing value contained in the Z-buffer are shaded. This prevents the order in which fragments are drawn from affecting the final image.

Transparent fragments generally do not use the Z-buffer, as elements in front and behind a transparent fragment will affect the rendered image. In order to produce a physically accurate final image, transparent fragments must be rendered back-to-front. A number of techniques can be used to overcome this limitation, but these are generally computationally expensive \citep{Everitt01}, 
or degrade the final image through stochastic sampling
\citep{End10}
to a point where a high number of overlapping fragments would become infeasible, either from an image quality or performance standpoint.

\begin{figure}[h!]
\vspace{-10pt}
\includegraphics[width=0.45\textwidth]{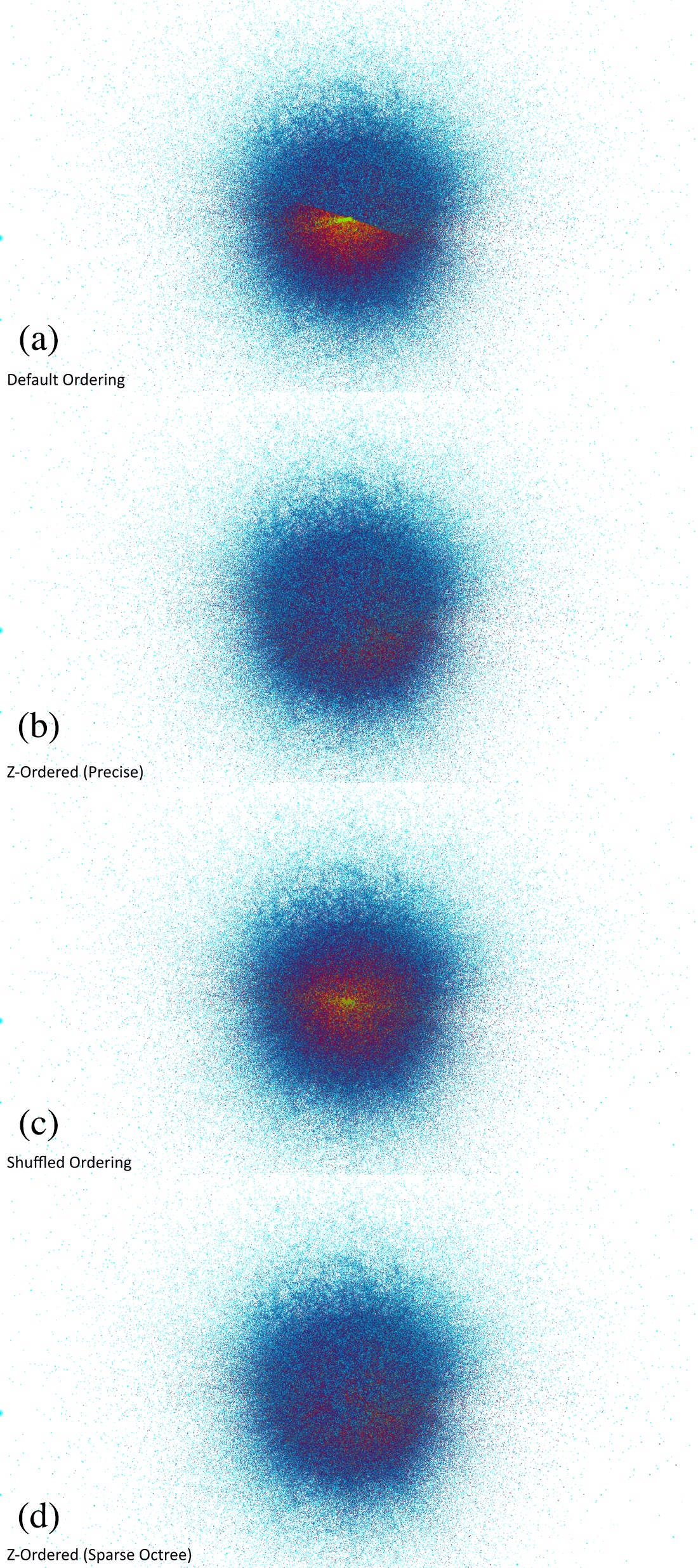}
\centering
\vspace{-5pt}
\caption{\footnotesize{
Effect of render ordering when rendering partially opaque particles. Four approaches are shown (top to bottom): (a) Default ordering, where the render ordering is defined by the order of entries in the data set; (b) Precise (ideal) ordering based on a fixed viewpoint; (c) Shuffled ordering, where the render ordering is randomised to remove bias; (d) Sparse octree ordering.
}}\label{fig:transparency}
\vspace{-15pt}
\end{figure}

The incorrect ordering becomes very apparent with highly transparent particles and certain viewing angles. Figure \ref{fig:transparency}(a) shows the effect when $\approx 10^6$ data points are rendered in the order in which they appear in the data table itself. Notice the systematic order errors around the origin, forming a sharp line. This is even more noticeable in VR, as the stereo rendering gives us a sense of depth, but the rendered ordering of particles does not match this depth. Figure \ref{fig:transparency}(b) shows the ideal render order, where the particles are sorted by decreasing distance from the user's viewpoint. However, this approach is not feasible for large datasets, as the following steps are required each time the user's viewpoint changes:

\begin{enumerate}
    \item Determine the distance squared from the camera to each data point (Complexity: $O(N)$).
    \item Sort the array of data points based on distance squared (Complexity: $O(N^2)$ worst case, more likely $O(N)$, as the order of points will only change slightly each frame).
    \item Update the GPU buffers with the new sorted array.
\end{enumerate}

 Our initial solution to this issue was to simply remove the systematic errors from ordering, by shuffling the data points, so they are drawn in a random order. In this case, every point is as likely to be in front of every other point as it is to be behind it, so the ordering between pairs of points is correct roughly half the time. In this case, systematic errors only crop up due to areas of high density being more likely to be rendered in front of areas of low density. This is shown in Figure \ref{fig:transparency}(c). The high density data points at the origin is still rendered in front of most of the radial data points. This is far less jarring in practice, and is also less noticeable when rendering particles with a lower opacity.
 
 An improved solution is to utilise an \textbf{octree} to partition the data set into a number of subsets (each consisting of a centre location, size and list of points in random order) and then apply a batch rendering approach to render them. The subsets can then be sorted by the distance from the camera to the centre of each subset, which applies an approximate ordering. As the sorting occurs on the the subsets, rather than the individual data points, this approach can be utilised in real time. Figure \ref{fig:transparency}(d) shows the results of this approach. The dataset has been divided into 32 partitions per dimension, for a total of 32768 partitions. However, as the data is sparsely distributed, only 5266 of these partitions are populated. The render time for the 5266 partitions was measured as 2.6 ms per eye, a slight increase from 2.0 ms when using an un-partitioned data set, while the inaccuracies in the final result are greatly reduced. One disadvantage of this approach is that the octree needs to be recomputed if the spatial mapping configuration changes. However, this is a minor issue for our application, as most datasets have a specific set of spatial coordinates (e.g. columns of Cartesian or spherical coordinates).   
 
In the IVL we have successfully rendered catalogues with 10$^7$ particles using a system with 32Gb of CPU RAM and a NVIDIA 1080ti-class GP;   see \S\,4.3 for example.  
 Larger catalogues are possible to render, but deleterious effects (noted above) are much more noticeable to the user.  We will be deploying a more efficient solution to overcome both large ($>$10$^7$) catalogues as well as dense (total particles per region) clusterings 
through the use of a hierarchical rendering system.

\subsection{Visualizing Volumetric Datasets}
\label{rendering:volume}
We now describe the \IDAVIE--v mode, central to our development in the IVL and the research efforts of the radio astronomy community.   
Image cubes in FITS format are read into system memory using the CFITSIO library, before being rendered in real-time using a GPU-accelerated ray-marching transfer function implementation adapted from the NVIDIA OpenGL SDK \footnote{\url{https://developer.download.nvidia.com/SDK/10.5/opengl/OpenGL_SDK_Guide.pdf}}. Ray-marching renders volumetric datasets by projecting rays from the user's viewpoint through each pixel of the screen, and ``marching''  each ray through the volume to determine the final colour of the pixel.
At each voxel intersection an accumulation calculation is performed, 
which affects the ray's value \citep{Levoy90, Vohl17}.. 
The two simplest examples of this are:
\begin{enumerate}
    \item \textbf{Average intensity projection (AIP)}: the ray's final value after propagating through the volume will be equal to the mean value of the voxels visited.
    \item \textbf{Maximum intensity projection (MIP)}: the ray's final value after propagating through the volume will be equal to the largest value of the voxels visited.
\end{enumerate}

Preliminary testing of these two methods revealed that the AIP method conveyed to the user a more distorted appearance, as if the data was seen through a warped glass surface, the amplitude of distortion dependent on the angle viewed.  In contrast, the MIP conveyed a more accurate representation of the 3D distribution, with little or no view-angle dependent distortion, and a level of transparency that allows easy visual inspection of the data.   Other transfer functions have been proposed and are likely to be very effective with spectral cubes  \citep[see e.g.][]{Vohl17}, but given the simplicity and (now) familiarity of the AIP/MIP transfer (e.g., this method is deployed in the popular astronomy application of SAOIMAGE-DS9), 
we have consequently adopted the MIP as the default
(although the user may switch to AIP if they desire). 

Moreover,  with our GPU-accelerated system, the user is able to change the intensity of the voxels, as well as the color transform, and thus highlight desired features, such as local maxima (astrophysical sources) and low surface  brightness `diffuse' emission.   In the current implementation of \IDAVIE--v, opacity is proportional to voxel intensity, which provides a visually satisfactory rendering of the cube.
We will implement fully-featured transfer function controls in future development, allowing the user to specify arbitrary color and opacity transfer functions, rather than those based on simple colourmaps.

\IDAVIE--v is designed to handle large image cubes, up to the size of system memory and beyond using virtual memory. However, there are two limitations that must be taken into account: Firstly, most machines have a smaller GPU memory capacity, so the cube must fit in GPU memory. Secondly, ray marching is highly sensitive to GPU memory bandwidth: For each rendered pixel, tens or hundreds of voxels from the dataset must be sampled. While many voxels will be sampled multiple times for a number of pixels' rays, the datasets are generally orders of magnitude larger than GPU cache capacity, so the cost of cache misses limit rendering performance. The second limitation is generally more stringent. We found that, even for high-end GPUs with over 10 GB of memory, cubes above approximately 1 GB in size were not able to be visualised within the target frame time of 11 to 13 ms (likely related to Unity limitations).  The solution to both of these issues is to use down-sampling, thus reducing the volume as needed. 

In order to effectively visualise larger cubes, we utilise a block downsampling algorithm to reduce cube size. The entire cube is read into RAM memory using a C++ plugin (making use of the CFITSIO library). We utilise a block downsampling algorithm in order to reduce cube size: Each $N \times M \times K$ block is reduced to a single voxel, where $N$, $M$ and $K$ are the width, height and depth of the block respectively. Any non-finite voxels in the original block are ignored during the downsampling process. We provide two downsampling strategies: the resultant voxel is either the mean value of the block's voxels, or the maximum value. The choice of this strategy is based on the user's choice of AIP or MIP accumulation in the ray-marching process\footnote{From our experience working with astrophysical volumes, such as HI spectral cubes, the MIP gives the most intuitive rendering experience for general ray-march rendering and for optimal downsampling.}.

The downsampling process is performed in parallel using OpenMP. The values of $N$, $M$ and $K$ can be specified by the user, or are calculated automatically in order to reduce a cube of size $W \times H \times D$ to the chosen maximum cube size. After downsampling, the reduced cube is uploaded to the GPU as a floating-point 3D texture, and then rendered on the GPU using custom shaders written in High Level Shader Language (HLSL). The cube is re-rendered at a frequency linked to the headset's refresh rate (usually 80 or 90 Hz).
The user is then able to explore the data (now rendered by the GPU), and choose specific sources or regions of the cube (see below), which may then be re-sampled to the original resolution (held in CPU memory). In this way we are able to work with large cubes and also maintain the native resolution for the most effective exploration.

A user can select a sub-section of the downsampled cube using the motion controllers. Once the user is happy with the selection, they can crop the cube to the selection. As the cropped cube has a reduced size, the values of $N$, $M$ or $K$ will be reduced, and the downsampled cube will more accurately represent the full-resolution data. If the cropped cube is small enough, the full-resolution data will be used without any downsampling. The user can instantly jump back to the downsampled cube and make a new selection. This process allows users to easily explore large cubes, while also preserving full-resolution and full-fidelity data in regions of interest. A sequence diagram of this process is shown in Figure \ref{fig:sequence} in \ref{appendixb}.  To date, we have successfully rendered cubes as large as 100 Gb using a machine with 64Gb RAM and 2080ti-class GPU.

In addition to loading the image cube, a mask cube of the same dimensions as the image cube (but in 16-bit integer format) can also be loaded. The mask cube can be used to selectively show or hide voxels that have been flagged by a source-finding software package such as SoFiA. The mask cube is cropped and downsampled in a similar manner to the image cube, and uploaded to the GPU. The GPU shaders then sample the mask cube as well as the image cube while ray-marching. This allows the user to instantly mask or unmask their rendered data.  Selecting individual objects or targeted regions generally means working with full resolution image and mask data, as is the primary method for source interaction.

Users can also edit an existing mask cube, or create a new one from scratch. By enabling ``paint mode", users can apply an additive or subtractive brush of a configurable size to the mask, thereby adding or removing voxels from the mask using their motion controller. When paint mode is enabled, an additional mask outline -- a wire mesh -- is generated and displayed. The mask outline is updated in real time. The outline is calculated based on scanning the mask cube and determining which faces of each voxel are bordered by a voxel with a different mask value. The ``active" faces for each voxel, along with the index of the voxel in the cube, are then encoded into a 32-bit integer. Voxels with no active faces are skipped; i.e., unnecessary segments of the wire mesh are snipped away automatically, thus providing a clearer view of the mask and the underlying data.
The list of encoded integers is then sent to the GPU, where a geometry shader unpacks the face information and creates a wireframe quadrilateral for each face in the list at the appropriate location (determined by the voxel index).  We demonstrate mask/voxel editing in Section 4.4.

\subsection{Cursors, Controllers and Voice Commands}

The hand controllers enable scaling (size), rotation and translation actions, allowing intuitive and natural access to the 3D data, menus and the space environment.   The controllers may be used to ``grab", select and and even move objects (e.g., menus), which is fully exploited within the \IDAVIE\ system.  Both controllers are used together to create additional motions (e.g., rotation is induced with a scissor-action from the controllers, and zooming in/out from moving the hands closer or further away from each other, etc).  And one controller may be used to hold a ``menu" (e.g., what we call the Quick Menu) and the other to select from that menu (i.e., your right hand pointing at your left hand in the VR to select ``painting" mode).   Hence both controllers are necessary for full functionality (but also see Voice Commands, below). 

Perhaps one of the most unique and powerful functions the controllers serve is to enable a 3D cursor.   Since we have full control of the space and the FITS data, the user may place a rectangular object, attached to their controller, onto a desired voxel or set of voxels.  This cursor is then used to target specific voxels.  Holding your cursor on a voxel, a small menu will appear next to your hand that shows the voxel index coordinates, floating point value and other other ancillary information that may be computed (e.g., the WCS coordinates,  the velocity or frequency, integer mask value, etc). In this way, the user may explore and interact with the spectral cube with a fidelity that extends down to a single voxel.   Later we will show examples of detailed interaction, one of which entails voxel editing, or simply ``painting", utilizing the controllers (buttons and motions), and the 3D cursor.

Finally, one of the most efficient methods for data interaction that is employed by \IDAVIE\ is the Voice Command.   A number of actions may be invoked by simply using a voice command to the built-in microphone on the \VR\ headset.   For example, instead of hunting around the menus to find the action that changes the intensity thresholds, the user may invoke the voice command ``edit min" to change the lower threshold:  vertically raising or lowering your arm is how you would change the threshold up or down in value, accordingly.   Haptic feedback (a slight buzz felt in your hand controller) alerts the user that the voice command has been accepted and executed.   We find that the voice command feature is extremely efficient -- much faster than hunting through menus -- and is generally robust in accuracy (voice commands are recognized with high repeatability).  The only downside to voice commands is remembering the actual command, which is a growing list.  The user has the ability (through a menu) to show a list of voice commands in case they are needed.

\subsection{Rendering Considerations for \VR}
\label{rendering:vr}
Maintaining a high, regular frame rate is essential for reducing motion sickness when using VR 
\citep{Weech19}.
\IDAVIE{} targets the default refresh rate of the headset being utilised (generally between 80 Hz and 90 Hz). In order to meet these targets, we apply some VR-specific optimisations. Firstly, SteamVR allows for a rendering resolution that scales automatically, depending on the user's hardware. This is activated by default, but does not respond to the changes in rendering time as the user changes their viewpoint. Secondly, we apply a custom fixed foveated-rendering\footnote{Optimizing the resolution quality by tracking where the eyes are looking,  enhancing central versus peripheral vision.} 
approach
\citep{Patney16},
which uses a larger number of ray marching steps within the small central area of the screen, decreasing the number of steps as a function of radial distance from the screen center. This improves performance at the expense of visual clarity at the peripheral. 

\begin{figure}[h!]
\includegraphics[width=0.50\textwidth]{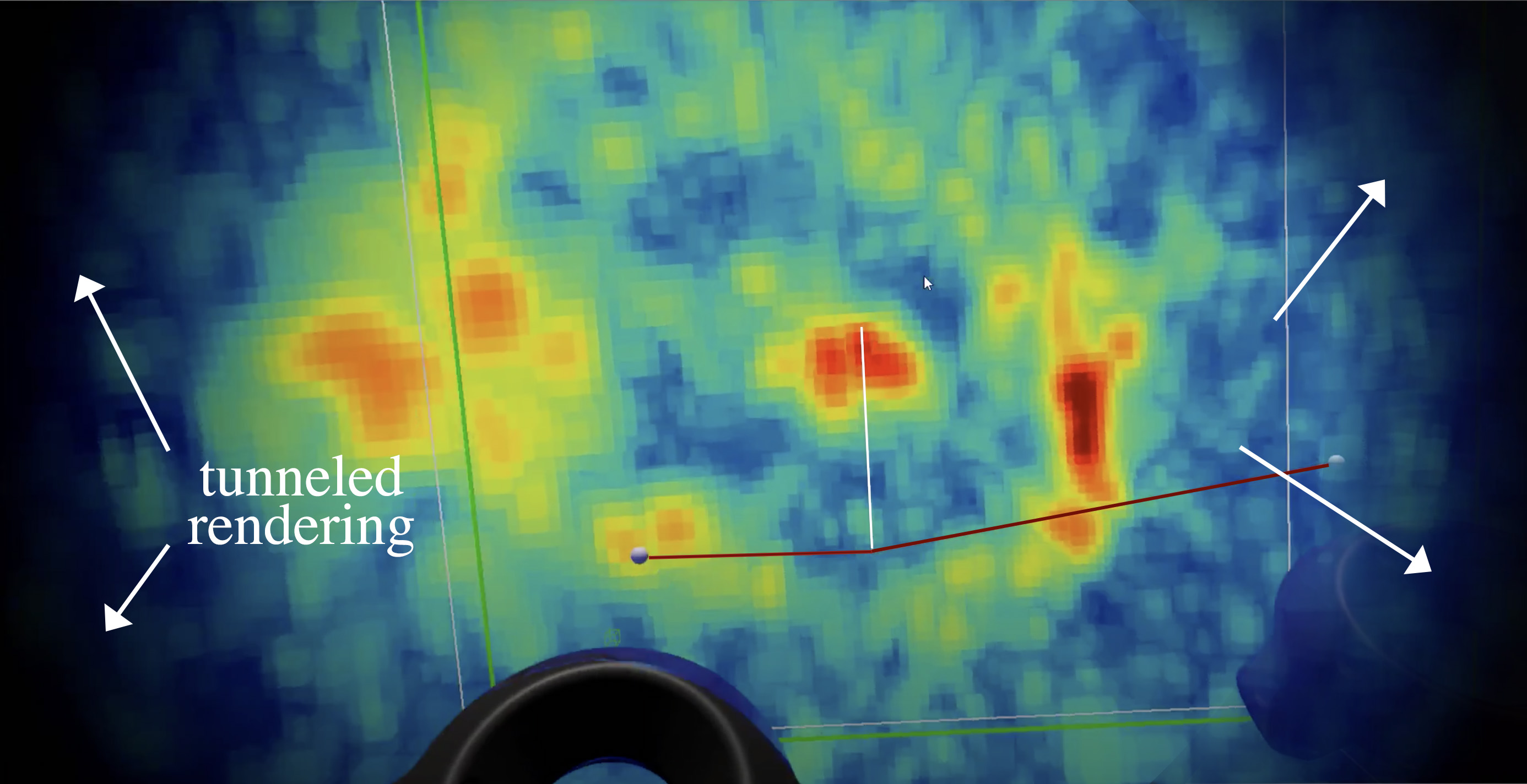}
\centering
\vspace{-20pt}
\caption{\footnotesize{
VR tunnel (vignette) rendering invoked during a hand motion (in this case, rotate and zoom).  Exploring a spectral cube with IDAVIE-v, hand controllers provide full 3-axis motion, as well as zoom in/out capability, all of which can induce vertigo if done too quickly.  
Peripheral vision vignetting of this kind helps to mitigate motions sickness, while foveated rendering helps to sharpen the central view.
}}\label{fig:tunnel}
\vspace{-10pt}
\end{figure}

A number of further adjustments can be made in order to improve the comfort of those users sensitive to motion sickness. One such approach, known as ``tunnelling'',
applies a vignette to the rendered image whenever there exists a disconnect between a user's physical motions and visually perceived motions, known to be a source of motion sickness in VR 
\citep{Kim18}.
The vignette decreases apparent motion in the user's peripheral vision, often decreasing discomfort. Unlike the traditional approach of applying the vignette uniformly across the rendered image, we apply a per-object vignetting, affecting only those objects that are not directly controlled by the user. For example, if exploring a volume dataset, the user's controllers will not be vignetted, as they are mapped directly to the user's physical motions. The user interface will not be vignetted either, as it is stationary with respect to the player's physical space. However, the dataset itself is vignetted when the user pans, zooms or rotates it; see example, Fig~\ref{fig:tunnel}.
Using this selective vignetting approach leads the user to feel as if they are handling a virtual object inside a physical space, rather than being teleported around a virtual space. 

\section{Science Results with \IDAVIE}

In this section we describe research projects that utilized the immersion technology developed in the IVL, and in particular the \VR\ system.  
These include interrogation of both particle and volumetric astrophysical data sets exploiting the \IDAVIE--p/t and \IDAVIE--v branches respectively.  We start with redshift catalogues (\S\,4.1), n-body galaxy-galaxy interaction simulations in the time domaine (\S\,4.2) and cosmological simulations of the local Universe (\S\,4.3).
The last two projects showcase, atomic hydrogen in the Andromeda and Fornax Cluster galaxies, which provide details of the \IDAVIE--v branch (sub-section \S\,4.4).

\subsection{All-Sky Redshift Survey Exploration}

One of the most natural applications of \VR\ with astronomical data are the 3D catalogues derived from large-area redshift surveys. It is difficult to separate and disentangle projection effects when viewing large scale structure (LSS) in 2D, regardless of whether it is rendered in 3D or in slices.  Conversely, in the full virtual reality environment, the user is able to move around the structures, either manually moving the data with the controllers, or simply walking around or within the data.  

\begin{figure}[h]
\vspace{-10pt}
\includegraphics[width=0.50\textwidth]{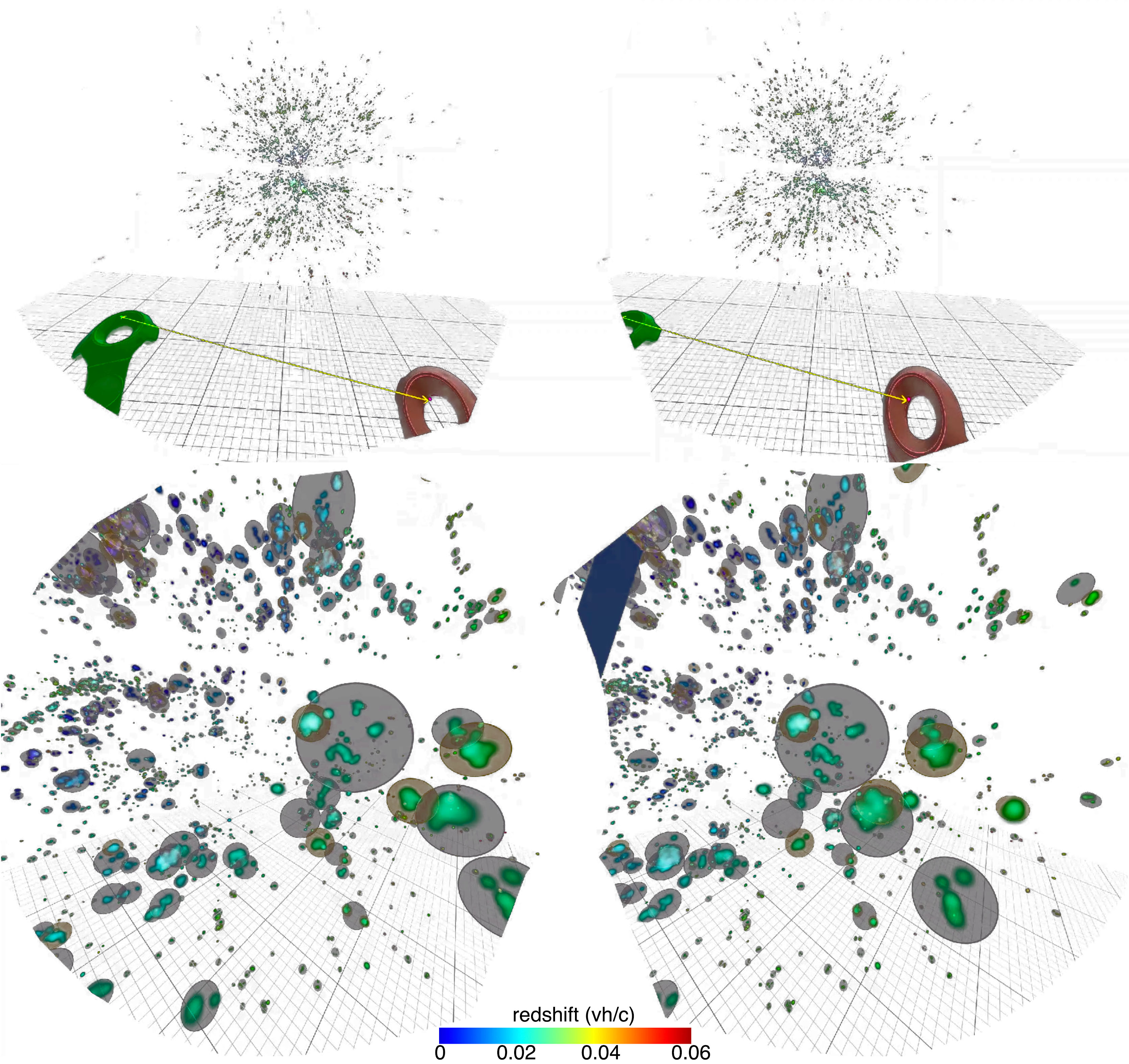}
\centering
\vspace{-15pt}
\caption{\footnotesize{
2MRS galaxies as viewed in \VR.  Here we show the left-eye and right-eye views. (upper) Extending to redshifts of 0.05 to 0.06, Over 45,000 galaxies rendered in cartesian X,Y,Z space. They are color-coded with a ``jet" rainbow transformation of the galaxy distance: blue is nearby, red is further away (note colorbar). (lower) zooming in, the dark grey are spheres that represent the galaxy groups of the new catalogue 
\citep{Lam2020}. 
}}\label{fig:2mrs}
\vspace{-5pt}
\end{figure}
The first application of this was carried out on the 2MRS \citep[2MASS Redshift Survey;][]{Macri2020},
 consisting of over 45,000 galaxies distributed across the whole sky with high-quality redshifts, and hence 3D positions, extending to redshifts of $\sim$0.06 (20,000 km/s). The 2MRS is (to date) the most complete picture of the `Cosmic Web' of galaxies that comprise the Local Universe 
 \citep{Bilicki2013}. 
The objective of the project 
 was to construct a new galaxy groups catalogue from the recently completed 2MRS.  Galaxy groups are gravitationally associated (and in most cases, bound) clusterings of galaxies, forming a unique environment (i.e., eco-system) within which galaxies form and evolve.  The burgeoning research field of Galaxy Evolution has determined that 
 the Cosmic Web is formed by the filamentary structures that consist of galaxy groups, containing almost half the mass of the local Universe \citep[e.g.,][]{Tempel2014}.
Hence determining group physical properties is crucial toward understanding galaxy growth and, in general, how they shape the evolutionary pathway of a galaxy. 

\begin{figure}
\includegraphics[width=0.49\textwidth]{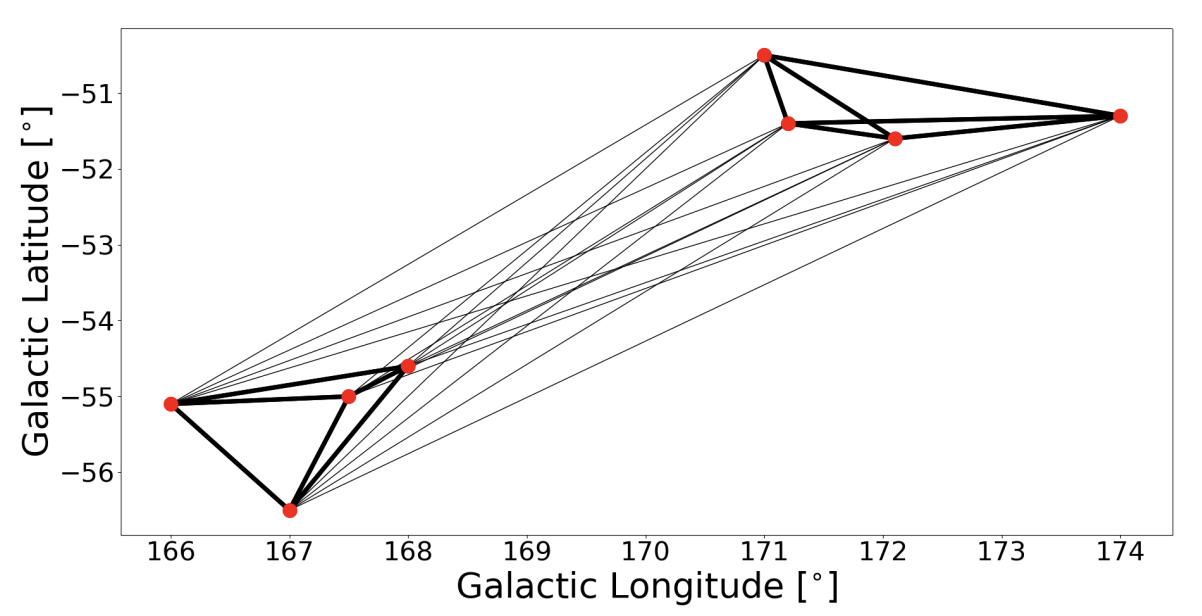}
\centering
\vspace{-15pt}
\caption{\footnotesize{
Graph depiction of a group of galaxies in close proximity to each other.   Points are galaxies, and the lines represent the strength of the grouping between any two pairs. Thicker edges represent strong bondings, while thin lines represent weak(er) bondings.  For the \VR\ system, instead of line widths, we use color to represent the pair weightings (see  Fig~\ref{fig:edges} ).
The graph is adapted from \cite{Lam2020}.
}}\label{fig:graph}
\vspace{-15pt}
\end{figure}

Described in \cite{Lam2020}
the groups catalogue (see Fig~\ref{fig:2mrs}, based on 2MRS redshifts and K-band (2.2 $\mu$m) luminosities, was constructed using a modified friends-of-friends (FoF) algorithm, with the novel approach to use graph theory as a visualisation aide for quality and assessment analysis; see Fig~\ref{fig:graph}.   The FoF employs a few critical parameters that sensitively determine which galaxies are associated with other nearby galaxies.  Complications arise from these simple prescriptions of spatial and velocity proximity that includes blendings (distinct groups in close proximity, appearing as one object) and chance alignment of galaxies along the radial (line-of-sight) direction.   Redshifts are fundamentally radial velocity shifts (from zero velocity to a higher velocity, based on the uniform expansion of universe).  Kinematic deviations from the Hubble Flow are called `peculiar velocities'.

Distances are crudely derived from these redshifts (using the Hubble Law and $\Lambda$CDM cosmology), which has the side-effect of inducing radial velocity biases in the distance.   The bias arises from the fact that galaxies are moving, relative to the space-time expansion due to local gravitational fields (e.g., the Milky Way and the Andromeda galaxy are moving towards each other because of their proximity and mutual gravitational attraction).   It is notably present when the galaxy belongs to a large group, or cluster.  The aggregate mass of a cluster can be quite large, thousands of galaxies ($>$10$^{14}$ M$_\odot$), which induces large relative velocities as the galaxies move and orbit within the gravitational well.  These peculiar velocities can range from 50 to 1000 km/s,  along the radial direction, which is of a magnitude that rivals the expansion velocities for nearby galaxies.  The result is to induce a radially stretched structure, sometimes whimsically referred to as ``fingers of god".   

We view radial-velocity `finger' structures in all redshift surveys.  They are both a nuisance -- because they are not representing the true locations in space and they encourage blendings -- and, conversely, they are flagposts or lighthouses of large clusters and groups in the cosmic web, making them easy to find.  In any event, group finding algorithms have to negotiate these radial fingers, disentangle and reconstruct the true distribution of galaxies\footnote{In addition to motion biases from individual galaxies, there are also bulk motions due to gravitational structures that are even larger than clusters, sometimes referred to as `attractors'}.  To understand and mitigate these velocity biases, a  new graph-theory visualisation based method was developed in the IVL \citep{Lam2020}.  It essentially draws a line between two galaxies (or ``friends") with a weighting  (represented with line weight/thickness, or a distinct color as presented below) that may be used to identify strong linkages (or friends) and weak ones, as well as sub-groupings (groups within groups); see Fig~\ref{fig:graph}.  

\begin{figure}
\includegraphics[width=0.49\textwidth]{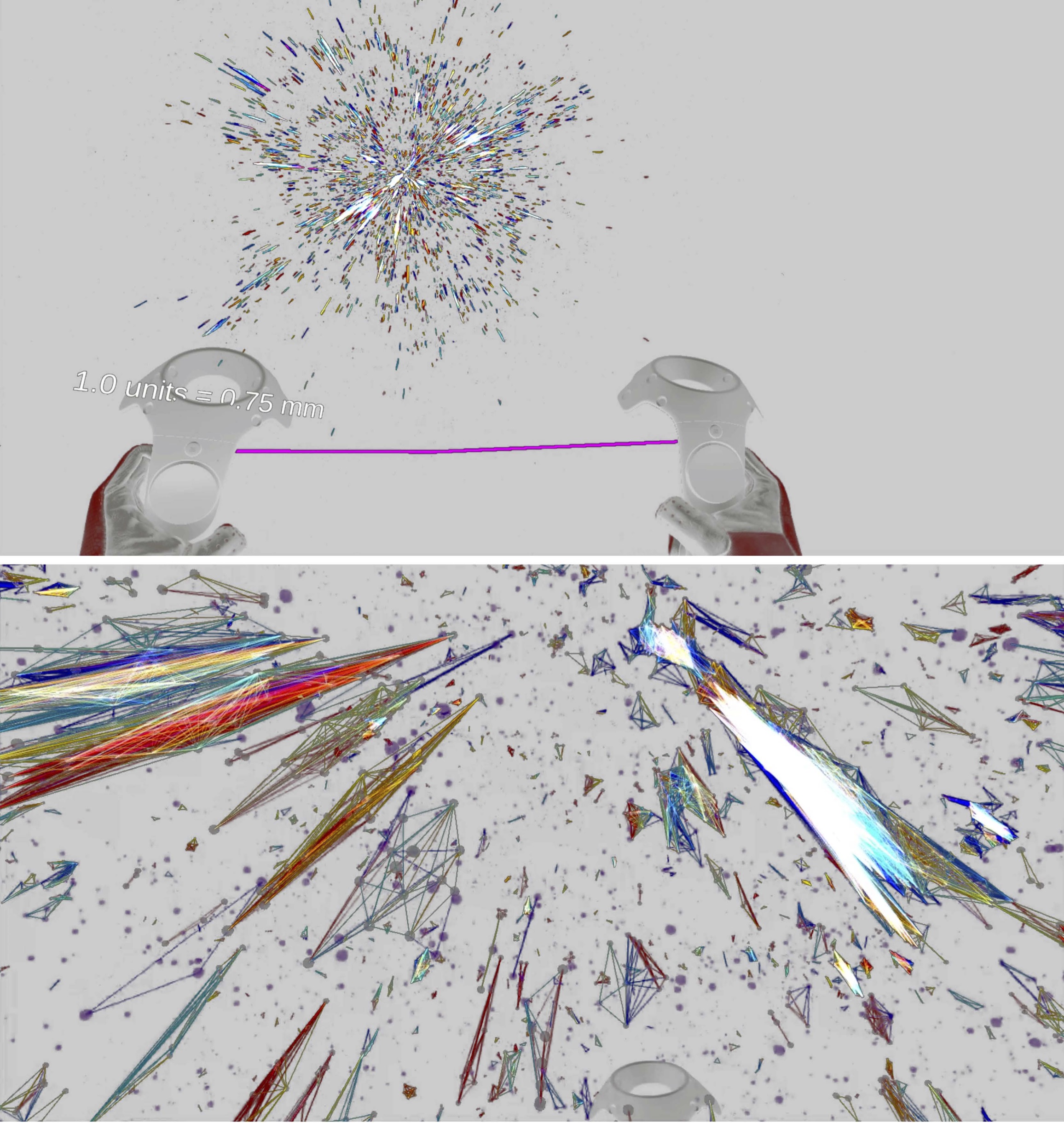}
\centering
\vspace{-10pt}
\caption{\footnotesize{
Visualizing the strength of galaxy groupings in the 2MRS using graph theory. Here lines are drawn between galaxies identified as ``friends" or gravitationally bound in groups.  The rainbow color coding indicates the weight or strength of the bonding, where red is strong, and blue is weak.  The color white is just an artifact of many lines blending together; in \VR\ we are able to zoom up the structures, move them around and clearly see the color weighting for small to large (massive) groupings.
}}\label{fig:edges}
\vspace{-15pt}
\end{figure}
In the \VR\ environment, we find that 
what renders clearly are thin lines assigned strongly contrasting hues; we used the common ``rainbow" transform to achieve contrast in these exploratory experiments, but we now regularly use the more optimal and improved ``Turbo-jet" transform, which is also available in the set of color transforms in \IDAVIE.   Fig~\ref{fig:edges} shows how the colour transform is used to represent ``thickness" or the weight of the galaxy-to-galaxy group connection.

Since there are many lines (45,000 galaxies, each may have dozens of friends, or lines), this may create a dizzy blur of lines.  Walking through them in \VR, zooming up regions and focusing on smaller areas, it is easy to see every line, disentangled and clear between pairs.  In this way we used the \VR\ system to analyze the preliminary results, adjust accordingly the FoF linking-length parameters, re-create the catalogues and repeat the analysis.  We found that blends along the radial (or z-) axis where common, and that weak linkages between neighboring groups or clusters was also a frequent.  Empirically we determined which pair-bonds (lines) to cut at thresholds that were robust to completeness and reliability balance. \textbf{Immersion visualisation,  in conjunction with graph theory,  was a crucial element toward constructing an all-sky galaxy groups catalogue that is both accurate and complete.}

\subsection{Galaxy-Galaxy Interaction}

Time evolving, numerical n-body simulations of tidal interactions between galaxies has proven to be a powerful way to understand how galaxies that are in close proximity, such as compact groups or galaxy pairs, evolve over relatively short times scales (i.e., measured in Myrs, as opposed to Gyrs).  Such tidal, or (at the most extreme) merging events have a profound effect on the angular momentum, gas distribution, star-formation and central-blackhole evolution for the individual and galaxy system.   The nature of these interactions is 4-dimensional, with a complex 3D spatial component that evolves over time in ways that are difficult (if not impossible) to intuit.  Computer simulations, followed with visualisation analysis to feedback upon the model parameters, is the only way to make progress in understanding this spectacular astrophysical phenomenon. We have found that \VR\ visualisation is ideal towards this end.

Here we highlight the `Fly-By' simulation
of two disk/spiral galaxies interacting and merging;
full details of the simulation and the most recent
version of the GalactICS code
can be found in 
\cite{Deg2019}.
In brief, 
GalactICS produces the initial conditions for the equilibrium models, including both collisionless components bulge, disk, and halo components, and a collisional gas disk.  The pair of galaxies in this simulation are initially identical, Milky Way type,
consisting of disk (younger and newly formed populations) stars, bulge/halo (spheroidal older population) stars, a gas disk, and a dark matter halo. Each galaxy has the same number of particles (i.e., total mass) and has fully formed stellar and gaseous
populations.  The pair start in some initial configuration, disks at some
angle to each with a particular parity (rotation of the disk). They are then evolved with 10 Myr time steps using the Gadget-2 code 
\citep{Springel05}.  At each time step, the state of each galaxy is recorded, including the positions of the particles. 

\begin{figure}[h!]
\includegraphics[width=0.49\textwidth]{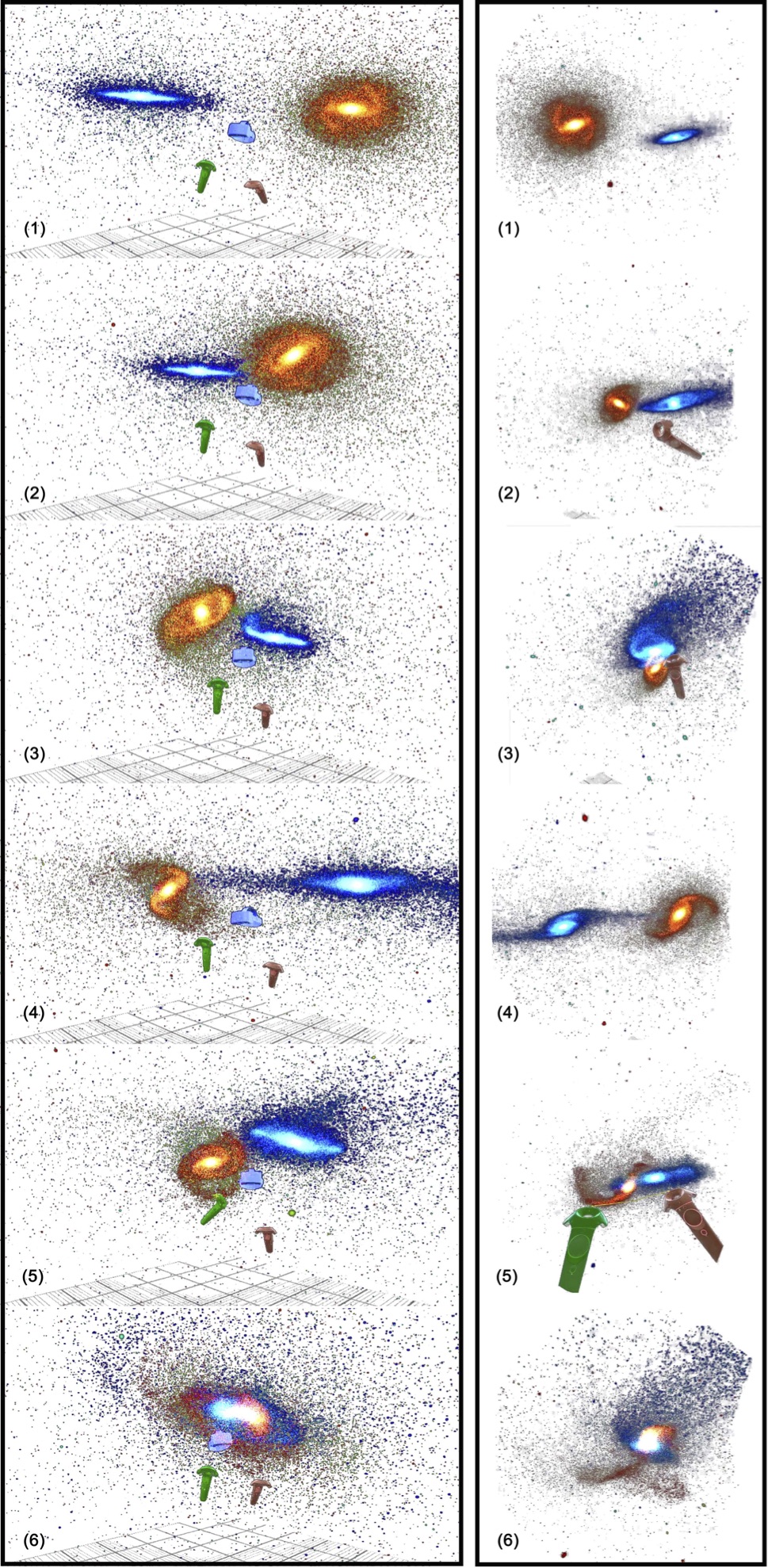}
\centering
\vspace{-20pt}
\caption{\footnotesize{
N-body simulation of an evolving tidal interaction between two Milky Way type galaxies of similar mass. Six states are shown (top to bottom) from initial separation to merged.   Left panels show the view from an external camera; right panels from the viewers left eye.
}}\label{fig:gg}
\vspace{-5pt}
\end{figure}

While the galaxies are initially identical, their orientation relative to the plane of the interaction is not.  As the simulation evolves, tidal forces disrupt the morphology of the galaxies, sometimes in non-intuitive ways.  For instance, in this particular simulation presented here, we start with identical galaxies in retrograde rotation (the disks are relatively counter-rotating) and inclined disk orientations with respect to the orbital axis.   After the first peri-centre passage, the two galaxies are no longer identical due to their difference in initial spin and spatial orientations.  With each passage, the pair become ever more distorted and disrupted, eventually merging into one system that is roughly spherical in shape. 

 Using our \VR\ system that is optimized for particles, we color coded each particle type: 4 for each galaxy, hence 8 particle types in total. 
The color coding was chosen so that a complimentary family of 4 color shades was assigned to each galaxy, making them both visually distinct and intuitive to the user as to which particles belong to which 
galaxy. One galaxy, e.g., would have blue assigned to disk stars, dark-green to halo stars, dark-red to gas and magenta to dark matter. The other galaxy would have orange assigned to disk stars, light-green to halo stars, burnt-orange to gas, and purple to dark matter.  An example of one 
particular simulation is presented in Fig~\ref{fig:gg}, demonstrating the contrasting color families between the two interacting galaxies.  The left panel shows the view from outside of the user, looking toward the simulation (the user is in the middle with the hand controllers and headset visible).  The right panels show the view from the user's left eye.  The initial state is the top panels, and the interaction evolves through the lower panels, until you see a merged system.  The user is taking careful note of the particle colors and locations throughout the simulation, with the ability to start, pause, and reverse the time steps; hence, with the \VR\ software, the user is able to see all angles (moving around , or manipulating the data orientation with the controllers) and all time steps, noting how particles are moving and congregating with non-linear tidal forces.

In general, the objective of these types of simulations is to produce merging sequences that are observed in nature, and to track where the particles have relocated. Some instances will result in compressed and enhanced gas, leading to star formation, and other instances where disruption ends any sort of star formation (i.e., quenching). Clearly it is a complex process. Analysis is carried out on each time state, computing surface densities and bulk flows (kinematics) of different populations (particle types). Standard visualisation uses plot and histograms, imaging and 3D rendering (in two-D slices). However, with VR it is possible to be inside the dynamic data set and track the evolution more naturally.  

Returning to the example simulation, the 
\textbf{difference in morphologies during intermediate time-steps is immediately apparent in VR visualisation, while it is more difficult to detect or even identify using the standard analysis tools.} 
Depth perception in a 3D view reduces confusion due to overlapping features, 
and thus the viewer is less likely to misconstrue the morphological information. This could lead to more accurate interpretations, and certainly less catastrophic failures (where the projection effect gives a completely misleading answer).  This is a unique and novel way to study dynamic simulations, and clearly demonstrates how new technology and visualisation methods -- namely \VR\ -- may be used to confront the shear complexity of interacting and merging extragalactic systems.


\subsection{Cosmological Simulations}



The Cosmic Web we observe today,  the large scale structure of galaxies in the local universe \citep[see][]{Jar04}, is the result of small perturbations (gravitational instabilities) in the 
large scale fabric of the early universe, evolving with time as galaxies form, grow and slowly die over the aeons.  At ever higher redshifts, astronomers observe the evolving cosmic web as brief snapshots in time, constrained by our instrumentation and our relatively short lifespans (compared to those of stars and galaxies).  
In the last couple of decades, a powerful new way to study the cosmic web that does not have these mortal constraints is through numerical simulations. Combining our best understanding of the physical laws of nature, cosmological theories or models (e.g., $\Lambda$CDM) and supercomputing resources, astronomers create simulations of the cosmic web from the early 
phase of the universe to the present (zero-redshift) epochs.  These simulations are multi-dimensional, including the standard 3D space and time dimensions; hence, lend themselves perfectly to advanced visualisation techniques.  In the IVL, we have been working with cosmologists to study their simulations using our \IDAVIE--p system.

\begin{figure}[h]
\includegraphics[width=0.48\textwidth]{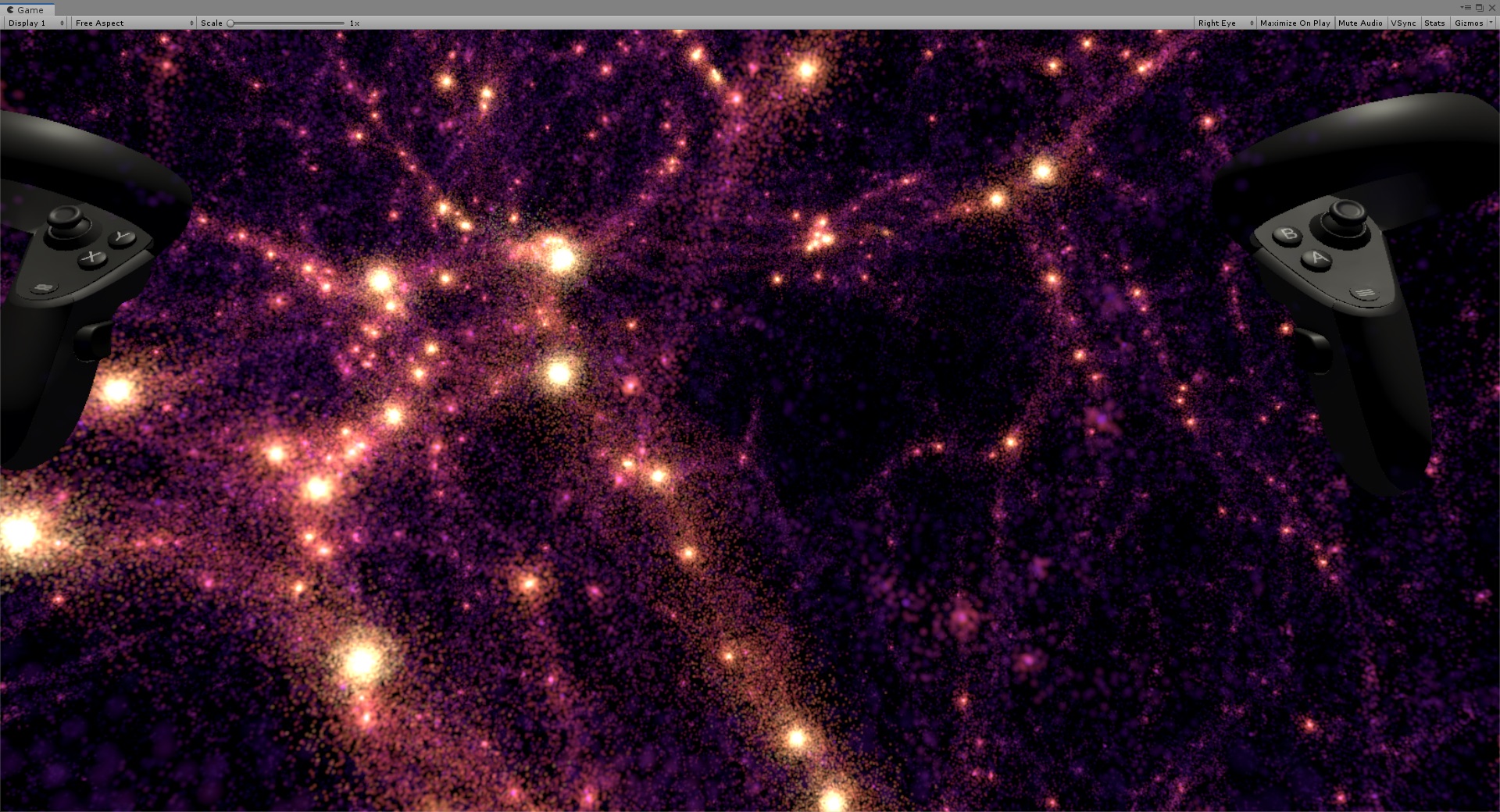}
\centering
\vspace{-20pt}
\caption{\footnotesize{\VR\ view of a cosmological simulation
of the local universe. Millions of gas particles are rendered using a ``plasma" color transform, where the brighter (yellow) signifies hotter temperatures.   Note the web-like filaments that connect to dense and hot `nodes', these are the massive galaxy clusters. Dissipative gas dynamics leads to the formation of systems we recognize as galaxies. Data and research from N. Katz (UMASS) and collaborators 
\citep{Huang2019}.
}}\label{fig:cosmo}
\vspace{-10pt}
\end{figure}

Here we focus on a simulation of a 50$^3$\,Mpc$^3$ piece of the present-day universe.  The simulations are described in
\cite{Huang2019}, who employ a
smoothed particle hydrodynamics (SPH) code on sub-grids that include pressure-entropy formulation,  time-dependent artificial viscosity, refined timestep criteria, and metal-line cooling.  The objective is to understand how mass is assembled (in and through the cosmic web) and galaxies are formed.  The most important physical mechanisms that control the baryonic processes are gas accretion, shock heating, and cooling, which are modelled with 
standard supernova-driven galactic winds
(and other feedback mechanisms).   A visualisation of the `universe-in-a-box' simulation is shown in
Fig~\ref{fig:cosmo},  depicted with a ``plasma" color-transform are hot gas particles with temperatures that range from 10$^4 -$ 10$^7$\,K.  


\begin{figure}[h!]
\includegraphics[width=0.48\textwidth]{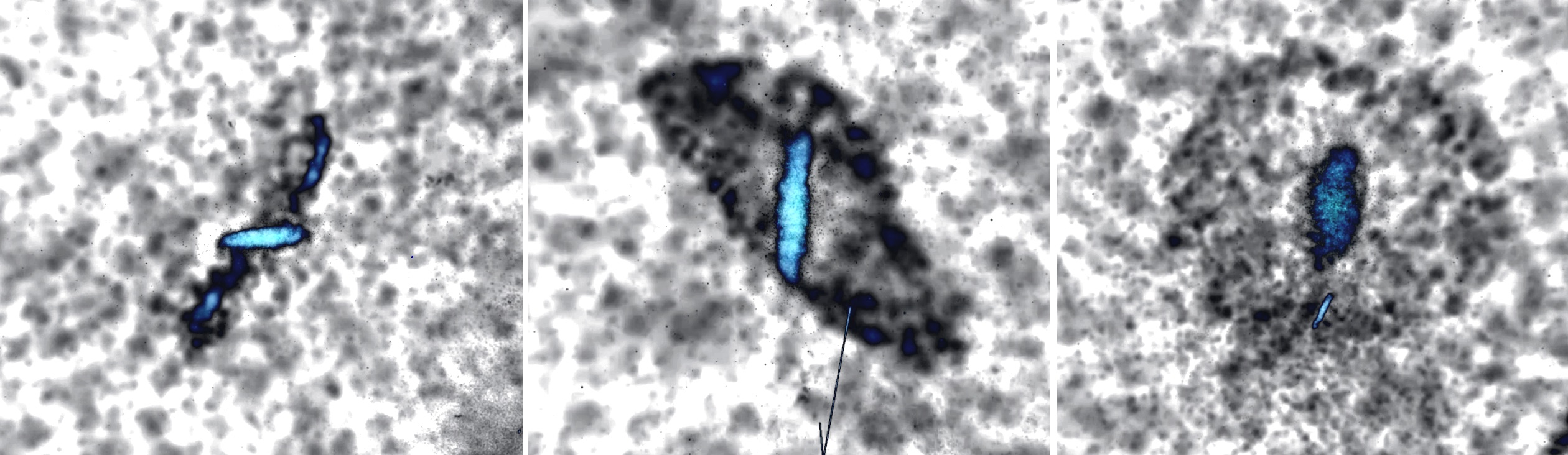}
\centering
\vspace{-20pt}
\caption{\footnotesize{Polar ring galaxy generated in the cosmological simulations, discovered using our \VR\ visualisation system.  Gas temperature is depicted with a grey-shade transform, and stellar ages are represented by a rainbow ``turbo-jet" from red (old) to blue (young stars).  Data from N. Katz (UMASS), who discovered the rare event using \IDAVIE--p.
}}\label{fig:polarring}
\vspace{-10pt}
\end{figure}
These new, high-fidelity cosmological simulations have angular (or particle) resolutions that are sub-kpc scales, which are good enough to see the internal components (structure) of galaxies themselves.  Using our immersive tools, we are able to explore the simulations, moving easily through large scale structure, ranging across temperature and density, at all scales from the entire 50$^3$ Mpc$^3$ cube down to individual galaxies with internal stellar populations that are easily discerned.  During the course of exploring the data using our \VR\ system, N. Katz discovered a type of galaxy that is rarely observed in nature.  This discovery would likely not have happened with traditional methods used for visualisation because of the multi-dimensional and confused (in 2D) nature of the data.

\begin{figure}[h!]
\includegraphics[width=0.48\textwidth]{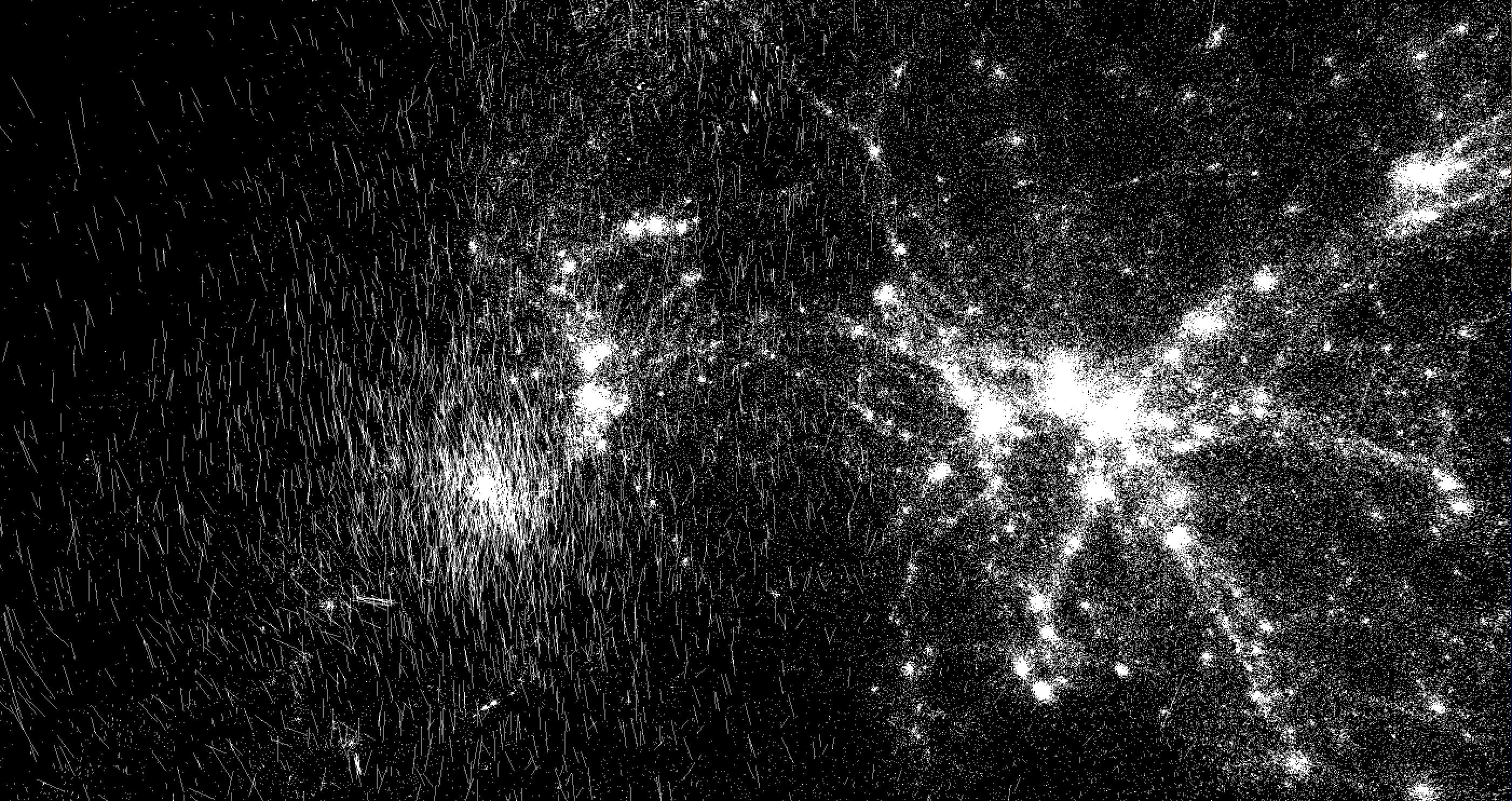}
\centering
\vspace{-15pt}
\caption{\footnotesize{
Gauging bulk flow motions in the cosmological simulation.  Here we use a B/W transform for the gas particles, and white vectors attached to each particle that indicates direction and speed.
}}\label{fig:cosmovec}
\vspace{-10pt}
\end{figure}

\begin{figure*}[b]
\includegraphics[width=1.02\textwidth]{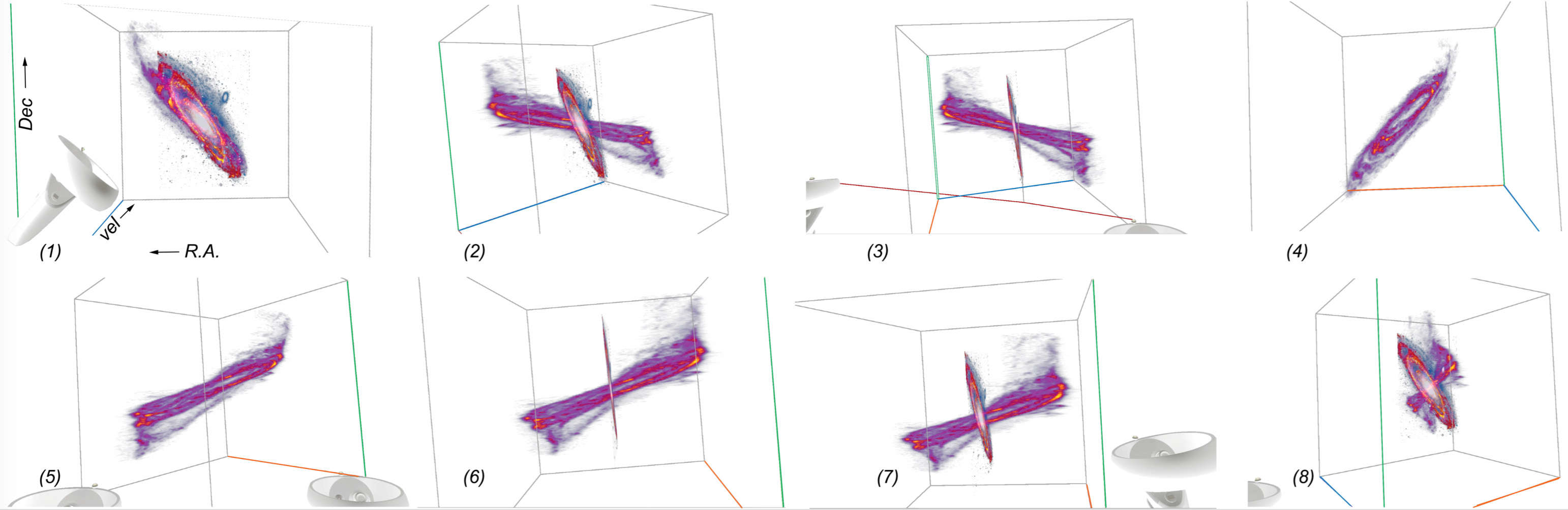}
\centering
\vspace{-15pt}
\caption{\footnotesize{
\IDAVIE--v view of the Andromeda Galaxy (M31). The sequence of panels (left to right) shows the galaxy at different viewing angles using \VR\ rendering of an \HI\  (atomic hydrogen) 3D spectral cube (X-Y-Z, where Z is velocity), revealing the extensive and fast (massive) rotation \HI\ disk.
Inserted at center is a more standard 2-D color composite image restricted to the spatial axis. The flat 2D image
disappears from view as we rotate around the spatial plane to see the velocity distribution of the gas. See also Fig~\ref{fig:m31snap} for a closer view
of panel (2).  Data courtesy of C. Carignan; \cite[see also][]{Carignan2006,Carig2}.
}}\label{fig:m31}
\vspace{-10pt}
\end{figure*}

Demonstrating the multi-parameter capabilities,
and now mixing both a grey-scale transform and a color ``turbo-jet" transform, 
we show both the gas particles and the stellar particles respectively, highlighting a striking object we identify as a ``polar ring" galaxy; see Fig~\ref{fig:polarring}.  This type of galaxy has the rare condition\footnote{Although these objects are rare in nature, they have even been observed in low density void environments \citep{Beygu13}}
of its star-forming, or gas, ``ring" acutely inclined with respect to the equatorial-flattened 
disk (i.e., quasi-polar alignment).  In the simulation, we have color-transformed (using rainbow) the stellar populations by age, where red are old stars and blue are young, newly formed stars; a different transformation, a family of grey shades, is used for the gas, to visually delineate from the star colors.  In this way, the \VR\ view clearly shows the orthogonal ring of gas relative to the disk of stars, whose relative ages are young (note the blue shading in the disk structure).  
These types of objects were predicted to be formed in simulations 
\citep{KatzRix1992}
but only now with 
sophisticated immersion tools can we truly explore the details of such rich simulations. 
{\bf The immersive and interactive environment permitted apprehension of the polar ring galaxy relatively quickly, essentially ``at-a- glance", which suggests a powerful synergy between the  volume-intensive cosmological computer simulations and the unique visual skills of human perception.}

Thus far we have viewed a time-static version of a simulated universe. They are dynamic, however, and constantly evolving. We have shown previously that multi-epoch views are possible in \VR, essentially toggling off/on adjacent epochs, controlling the time component while viewing the spatial distribution.  Another effective visualisation technique is to depict motion (or time-evolution) using a vector arrow.  The vectors have direction and amplitude (speed) that indicate the flow of motion, giving some idea of the time evolution of the structure.  This method is only effective for small time scales during which the vectors themselves only minimally evolve.  An example is shown in Fig~\ref{fig:cosmovec}.  In future development, we will be exploring graphical methods that better visualize vectors and flow lines, methods that have been very effective with 2D maps in a number of fields, notably meterological (e.g., wind flow maps).
There are many applications in astrophysics that use vector fields and flow lines in 3D scenaries, including polarimetry and magnetism,  large scale structure and cosmological simulations, of which \VR\ will be a powerful data interaction pathway.

\subsection{Volumetric Data Exploration:  Working with Cubes in \VR}

In this final sub-section we describe and demonstrate the \IDAVIE--v system performance with astrophysics science projects conducted with our collaborators and students.  
This is the most developed and sophisticated system in the IVL activity, and includes a public release of a beta version of the software \citep{angus_comrie_2021_4614116}
{\footnote{https://idavie.readthedocs.io/en/latest/}.

Two examples are showcased below.  The first is a common application where the user wants to simply view the spectral (spatial+kinematic) cube in the context of the more traditional 2D imaging of spatial and intensity information.   The second example is a more detailed dive into cubes, exploiting the full power of the \IDAVIE--v system.

\vspace{+5pt}
\underline{1.  Example of 2D map + 3D cube of the  M31 Galaxy:} \\

Here we show the spatial-kinematic distribution of the neutral hydrogen gas that fuels the growth and evolution of the magnificent Andromeda Galaxy. With our \VR\ system, we are able to freely explore the largest galaxy (angular-wise) outside of the Milky Way itself, whose physical size and mass are twins to that of our home Galaxy 
\citep[see ][]{Jar2019}.   With large and complex galaxies such as M31, 
it can be daunting at times to know your location bearings when in the \VR\ environment.
Spectral cubes are not spatially 3D, rather the 3rd dimension is velocity or frequency of the emission line.  We find it helpful to place an optical or infrared 2D image within the cube as a sort of spatial roadmap to since astronomers are used to looking at flat, spatial projections in these wavelength windows.

\begin{figure}[h]
\includegraphics[width=0.49\textwidth]{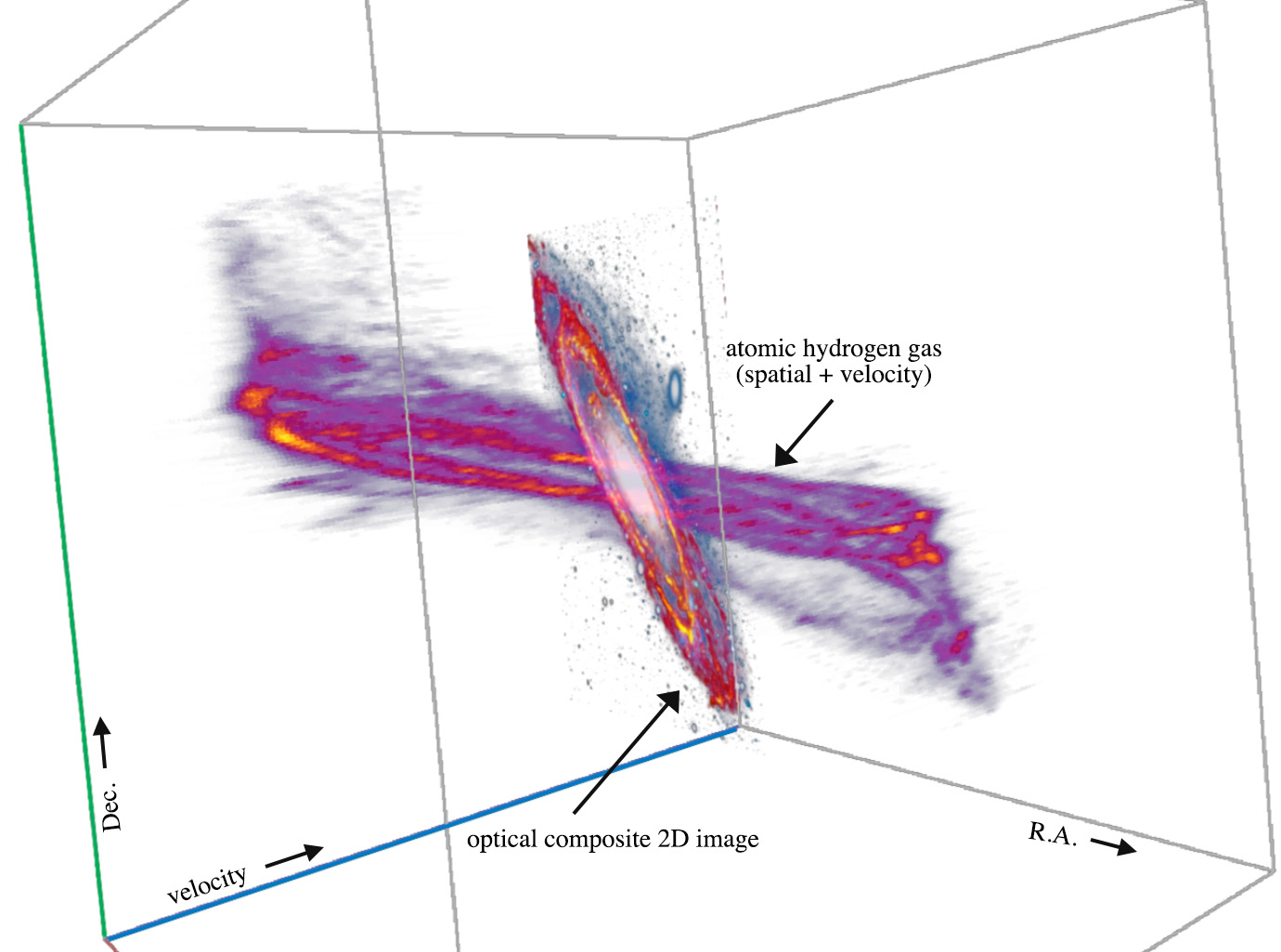}
\centering
\vspace{-5pt}
\caption{\footnotesize{Single snapshot from the M31 sequence (Fig.~\ref{fig:m31}), showing more detail of the 3D \HI\ gas distribution (spatial + velocity) and an optical composite 2D image inserted into the \IDAVIE--v space.  This is an example of how multi-wavelength images and volumetric cubes may be combined visually.  
}}\label{fig:m31snap}
\vspace{-5pt}
\end{figure}

Moreover, opt/ir imaging is sensitive to the sites of active star-formation, and hence the combination of HI (fuel) and imaging (fire) reveal the past and present growth of the system.
Our development \VR\ system allows for this dual-input functionality, with the requirement that the 2D map matches the spatial field of view of the cube, while our beta-release \citep{angus_comrie_2021_4614116} of \IDAVIE\ does not yet have the ability to insert JPEG or FITS images
(future versions of \IDAVIE\ will use the WCS information to properly match FITS 2D images with the 3D spectral cube).  An example is shown in Fig~\ref{fig:m31}, the \VR\ view (from the user's left eye) of the M31 atomic hydrogen distribution from a sequence of viewing angles \citep[data courtesy of C. Carignan; see also][]{Carignan2006,Carig2}.  To better discern details in this rendering,  a closer inspection of one of the views is shown in Fig~\ref{fig:m31snap}.

Since the star-forming disk of M31\footnote{See also the exquisite infrared imaging from the IRAS, Spitzer, WISE and Herschel space telescopes, showing the SF regions with pc-scale detail in M31; e.g. \citet{Jar2019}}
containing the \HI\ gas is rotating, the line-of-sight components of the gas are offset with respect to the systemic velocity of the system.  We refer to these offsets as ``blue" and ``red" shifts, where blue is negative relative to systemic, and red is positive relative to systemic.  We clearly see the rotating gas, which is aligned to the spiral arms seen in the 2D optical image  (which in turn is sensitive to the star-forming disk, spiral arms and stellar associations), combining the best of both scientific visualisations.   In addition, \IDAVIE--v is able to track the gas distribution and the kinematics through the 2D moment-0 and moment-1 maps (showcased below), which compare directly with 2D photometric imaging.

\vspace{+5pt}
\underline{
2. Neutral Hydrogen Content in the Fornax Galaxy Cluster:} \\

The \IDAVIE\ system was designed and developed with user interaction foremost in mind.  Velocity cubes have particular requirements, with the most desired functionality being:  
\begin{itemize}
\item voice commands to control intensity, color transforms, and Z-axis scaling adjustments, as well as other simple interactions such as display of critical source information; 
\item working with input source tables;
\item overlaying source masks on the data cube voxels; 
\item editing source masks in real time;
\item derive in-the-fly sky and source statistics, moment maps and other analytics that may be derived from the data and masks.
\end{itemize}

Here we demonstrate these features using a \HI\ spectral imaging of the Fornax Galaxy Cluster,  acquired with the Australian Compact Telescope Array (ATCA; data and ancillary information courtesy of P. Serra and collaborators; see
\citealp{Serra2016,Maddox2019}) and recent science results presented in \citet{Kleiner2021} that utilized the \IDAVIE\ system for analysis.
Fig~\ref{fig:fornax} is a \VR\ view of the \HI\ data cube, revealing a relatively large mosaic of over 30 deg$^2$,  covering the redshift range of the cluster ($\sim$2000 km/s). Fornax is graced by a spectacular starburst galaxy, NGC\,1365, which is gas rich, and exhibits a massive stellar bar, older bulge population, and a fast rotating disk \citep[cf. ][]{Jar2019}.
It will be featured in the graphics to follow.

\begin{figure}[h]
\includegraphics[width=0.49\textwidth]{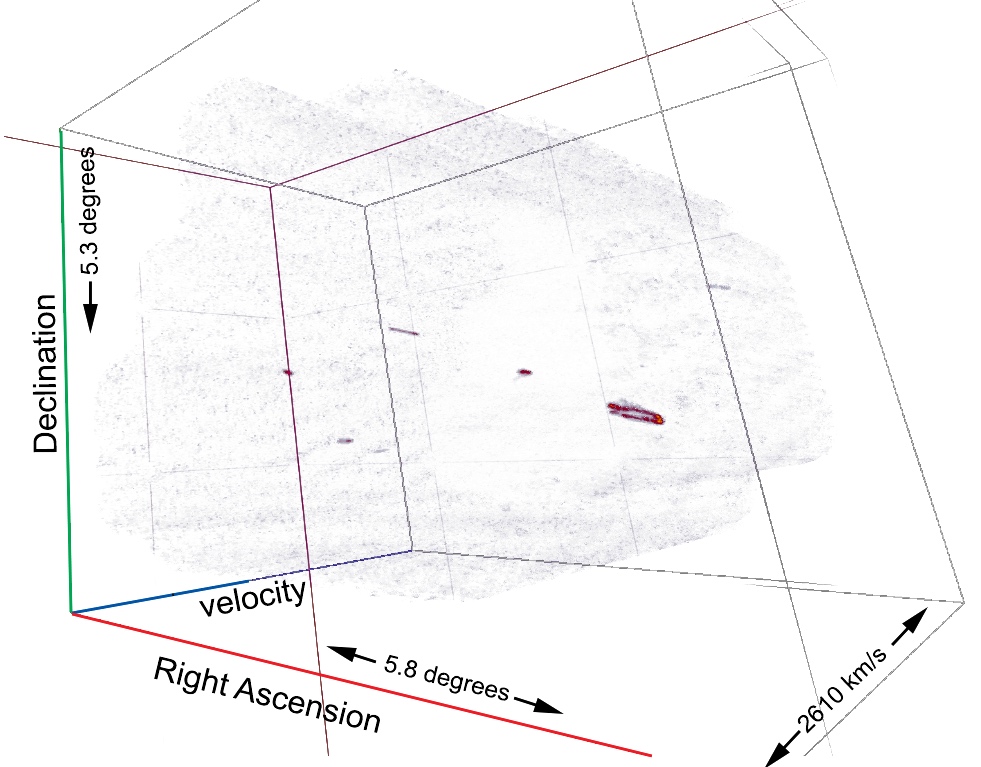}
\centering
\vspace{-15pt}
\caption{\footnotesize{
The Fornax Galaxy Cluster \HI\ spectral cube, showing the central region and velocity range of the main cluster as viewed in our \VR\ system. The data are derived from ATCA, and are courtesy of P. Serra.
}}\label{fig:fornax}
\vspace{-5pt}
\end{figure}

Source finding, extraction and characterization is a major step with \HI\ surveys.  Recently developed for the SKA pathfinders, the SoFiA\footnote{https://github.com/SoFiA-Admin/SoFiA; \citep{Serra15}} software pipeline carries out these steps in an automated fashion (with plenty of parameter tuning knobs to work with).  SOFIA creates a source list table (ascii or VOT) and a mask that is a FITS cube of the same size as the input data cube.  The mask consists of voxels that have been turned ``on" or tagged with the identity of the source.  In this way the mask tracks which voxels are associated with identified sources.
\IDAVIE--v has incorporated these SoFiA data products into the system.  The user is able to interact with the mask and the VOT source table, using the \VR\ to visually validate (completeness and reliability) the performance of SoFiA, as well as the ability to modify the source parameters through editing of the mask (see below).  A simple voice command: ``mask on", activates the mask -- which includes sending haptic feedback to the controller letting the user know the command was recognized -- and so the user sees only the masked voxels (the other voxels are turned off), as demonstrated in Fig.~\ref{fig:fornax1a}.

\begin{figure}[h]
\includegraphics[width=0.50\textwidth]{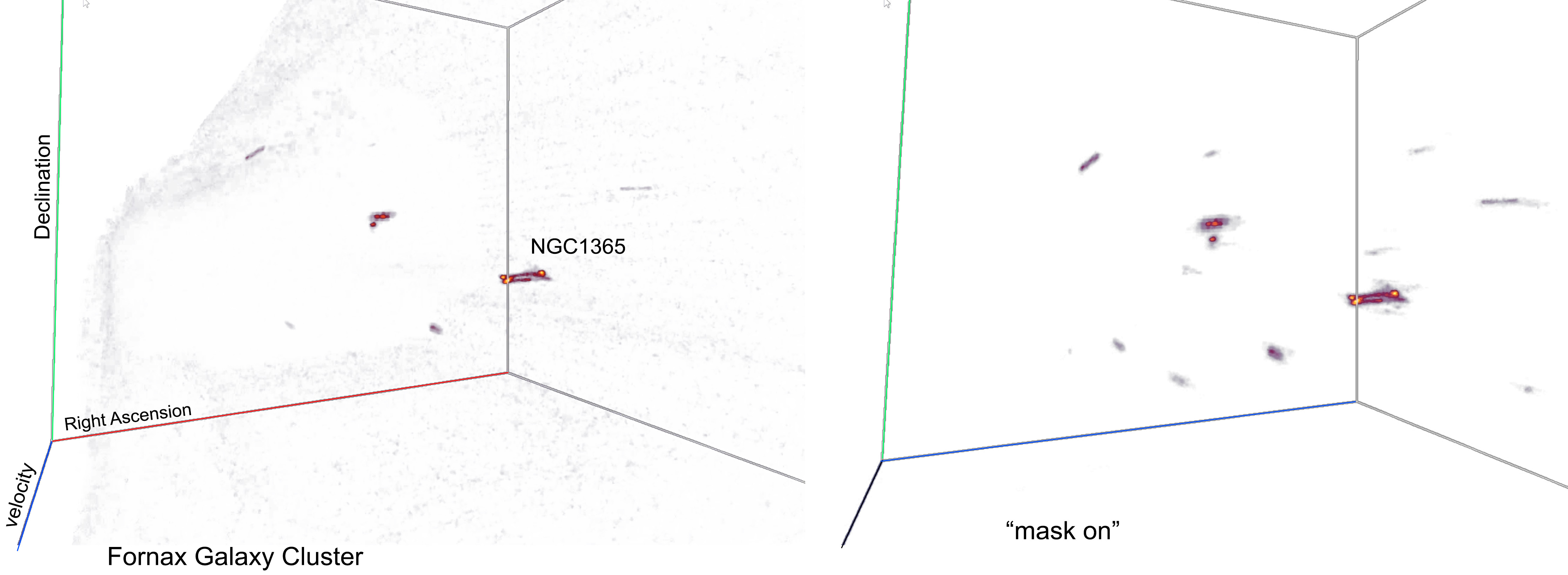}
\centering
\vspace{-5pt}
\caption{\footnotesize{
Front \VR\ view of the Fornax Galaxy Cluster.  The galaxies are not as easily discerned in the left panel, but clearly seen when the mask is turned on (right).  Sources that have been identified with their associated extended emission are visible, and all other voxels are turned off.
}}\label{fig:fornax1a}
\vspace{-5pt}
\end{figure}

The masks may be used in other useful modes as well.  Invoking the voice command ``mask isolate" will turn on the masked voxels, while everything else is turned off. This gives the highest contrast possible between what may be real astrophysical signal and what is simply noise.  An example of this mode is shown in Fig.~\ref{fig:fornax1b} (left panel). The final mode that masks may be invoked 
is 
``mask invert", which  turns off all voxels that are associated with the sources; see  Fig.~\ref{fig:fornax1b} (right panel).  In this mode, the data cube is stripped of identified sources and their associated extended \HI\ emission, and what is left is noise.  This is useful when looking for faint extended emission that the source finder missed, or if the user desires to look for new sources, which may be fainter and more compact and beyond what the source finder was tuned to seek out.


\begin{figure}[h!]
\includegraphics[width=0.50\textwidth]{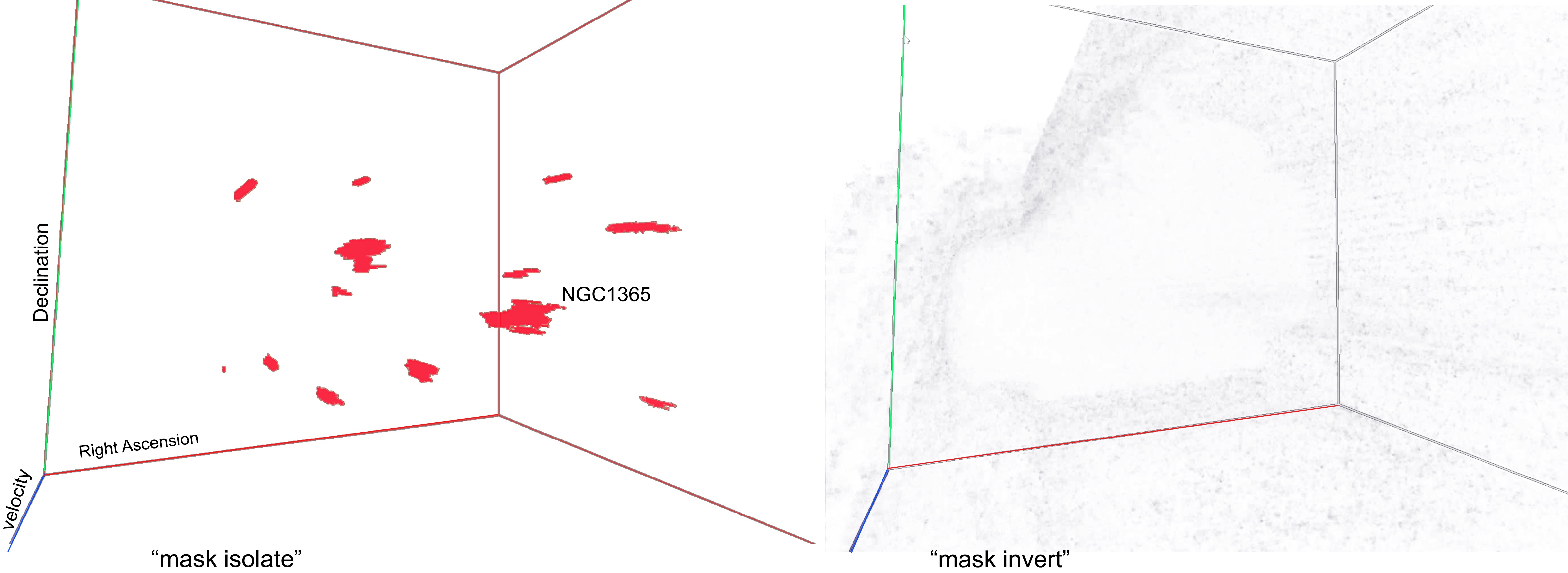}
\centering
\vspace{-15pt}
\caption{\footnotesize{
The Fornax source mask is now ``isolated" (left) to enhance the voxels that have measured \HI\ emission. Conversely, these masked voxels (seen in red) that are identified astrophysical sources, may be turned off completely -- voice command ``mask invert"  -- thus leaving the cube with only noise (right panel). In this way new sources may be found, or sources with associated faint emission that was missed by the source finder.
}}\label{fig:fornax1b}
\vspace{-10pt}
\end{figure}

Resolved and extended sources may have complex distributions, both spatially and kinematically.  For example, the warps and off-planar gas (possibly associated with accretion processes) may appear asymmetric and kinematically offset from the rotating gas disk 
\citep[see ][]{Lucero2015,Heald2016}; 
hence, the source finders may have trouble identifying or associating it with its parent host galaxy. The solution is to deploy the powers of human pattern recognition -- edit the mask in the \VR\ environment where all of the information is at hand in the most natural (that is, 3D) environment to work. 

\begin{figure}[t]
\includegraphics[width=0.40\textwidth]{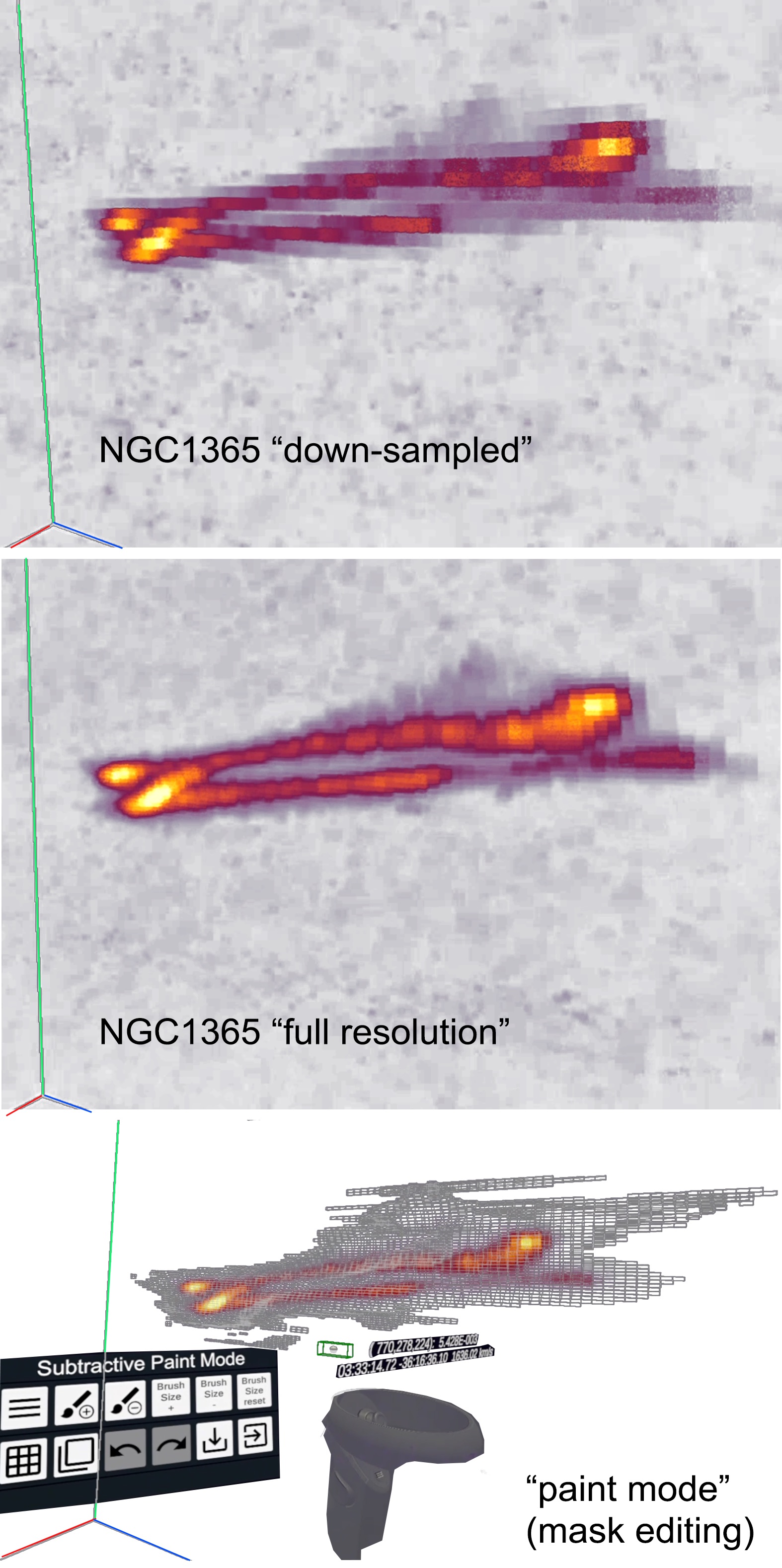}
\centering
\vspace{-10pt}
\caption{\footnotesize{
Editing the mask of NGC\,1365 using the ``paint mode" of the \IDAVIE--v system. Top panel shows the source in the down-sampled cube; middle shows the native resolution after it has been selected and restored to native (full resolution) condition; bottom panel the paint mode has been invoked, with subtractive mode (removing or erasing masked voxels), and the paint brush adjusted (3D cursor:  note the green box, and ball in center).
}}\label{fig:fornaxpaint}
\vspace{-10pt}
\end{figure}

Paint Mode:   Mask (or voxel) editing is incorporated into the \IDAVIE--v design, and is implemented in the mode that is called ``mask painting".  First the user selects the source from the down-sampled cube; with hand-controller selection, the source is re-displayed at full (native) resolution.
The user enters paint/edit mode through a menu that is accessed in the \VR\ environment through the hand controller or through a voice command.   The painting features include: changing the paintbrush (i.e., 3D cursor) size, and either subtracting or adding to the mask.

In paint mode, the user sees the mask of the source that has been selected, it appears as a mesh grid (the voxel intensities are still visible within the mesh).   Selecting the add or subtract mode and adjusting the brush size, the user is then able to ``paint" or erase mask voxels using the controller hand motion.  Visually discerning which voxels need editing, the user modifies the mask accordingly.  When completed, the user is able to save the modified mask as a new cube file.  An example of the paint mode is shown in Fig.~\ref{fig:fornaxpaint}, featuring the \HI\ gas and velocity distribution of NGC\,1365 and its associated mask (mesh grid).   {\bf The voxel ``paint" editing is one of the most powerful features of \IDAVIE\ because of the unique ability to manipulate 3D data in a natural environment, free of obstruction and blending that is an unfortunate property of 2D and pseudo-3D tools.}

Moment Maps:   As part of the beta-release of \IDAVIE--v, moment maps are constructed in real time using GPU shaders, with  the results displayed to a dedicated graphics window in the \VR\ environment.
An example is shown here for NGC\,1365, Fig.~\ref{fig:moments}, where the moment-0 and the moment-1 reveal the 2D spatial gas distribution and rotating kinematics of this large barred spiral galaxy.  This is designed to work best with single objects and their masked ``on" values, selected by the user, which allows for real-time fast and efficient GPU computation and CPU rendering.  The user 
has the ability to change the thresholding of the rendered maps to enhance contrast.


\begin{figure}[h!]
\includegraphics[width=0.49\textwidth]{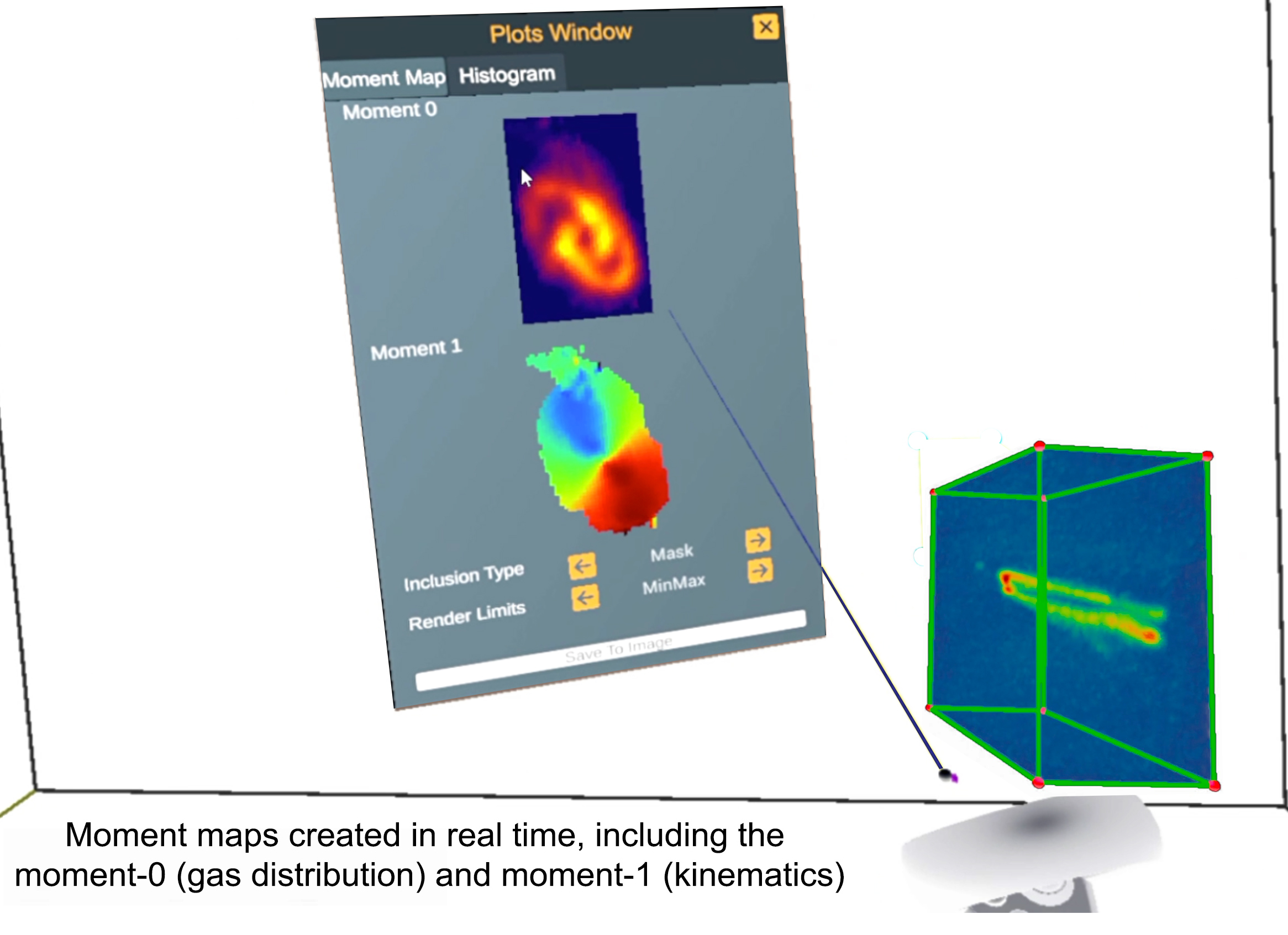}
\centering
\vspace{-25pt}
\caption{\footnotesize{
Moment maps are rendered in real time and displayed in the \VR\ graphics window, including the moment-0 (2D gas distribution) and the moment-1 (radial kinematics).  Here we show the resulting maps for the \HI\ gas of NGC\,1365, which has been ``crop selected" within the spectral cube (see right side).
}}\label{fig:moments}
\vspace{-10pt}
\end{figure}

 By default the moment-0 is rendered with a ``heat" colourmap,
which nicely conveys,  through a sequential colorbar,  monotonically increasing (perceptual) lightness (i.e., increasing intensity of the \HI\ gas, which traces gas mass distribution).
The moment-1 is rendered specifically with the  
``turbo-jet" colourmap, which has the useful property in being 
two-sided in a quasi- symmetric way in hues;  this colour map 
visually and intuitively conveys  the blue shifted gas (relative to systemic) and the red-shifted gas (relative to systemic), with the systemic gas (low rotation velocities) appearing with green/yellow hues.
For rotating disk galaxies, you achieve the characteristic rendering as shown in Fig.~\ref{fig:moments}, the starburst galaxy NGC\,1365.\\

Finally, it bears noting that \IDAVIE--v has the capability to upload tables with redshift information (i.e., spatial coordinates and the vh/c redshift) which may be overlayed and viewed within the cube.  In this way the user may compare the 3D spectral distribution with that coming from archival catalogues.  See the beta release documentation, link provided below in Section 5.1.
Functionality that leverage developed applications from the \IDAVIE--p  (particles) system, such as catalogue manipulation,  will be more prevalent with future development of the volumetric \IDAVIE--v system.


\section{Discussion and the Way Forward}
In this last section, we discuss some of the lessons learned, challenges we continue to face and bright prospects looking forward.  We begin with the genesis of the IVL and \IDAVIE\ development, our R\&D experiences over the last three years, the recents software release,
and what we see as the game-changing developments that are to come in the next few years.

\vspace{+10pt}

Facing the prospect of doing science in the Big Data era of the SKA and other major next generation telescopes and surveys (notably LSST), we created the IVL to fully exploit immersive technology.  Five guiding principles toward this end were:   
(1) multi-disciplinary flexibility, 
(2) collaborative platforms,
(3) transformative and innovative,
(4) cloud-based interfacing, and 
(5) advocacy and engagement with our peers 
and the interested public.  Our visualisation lab deployed large format ``wall" screens, curved and full-domed facilities, and finally \VR\ stations.  

The promise of virtual reality is well known:   most suited to multi-dimensional data sets that includes particles and volumes, it combines discovery and (through software tools) quantitative analytics, 
it is the most 
intuitive way to interact with data.
3D rendering enables far more 
information that can be conveyed in comparison to standard 2D,
and opens up new methods for 
 analysis and data
exploration.  It is the last point that has been barely tapped, and whose potential is nearly inexhaustible.  But it is also quite challenging, optimizing code for multi-thread GPU programming and working within the confines of the rendering (``gaming") engines, not to mention the bespoke equipment and underlying system software that is required.   

\subsection{Data and Science Driven Initiatives}

In this paper we have highlighted the work that has been carried out in the IVL over a span of about 3+ years (2017-2020, and now continuing into 2021).  The development was driven by scientific research that our colleagues and international collaborators are conducting, focusing on those that are notably 3D (or more dimensions) in natural form.  
In most cases, we worked closely with the researchers to create interactive tools and functionality that were applicable and advantageous to their data analysis. 
 Some of these tools were directly inspired by well established 2D applications  -- the powerful SAO-DS9 viewer favoured by opt/ir astronomers -- the equally powerful KARMA\footnote{https://www.atnf.csiro.au/computing/software/karma/} suite (especially KVIS) for radio astronomers, and other 2D and 3D
software that are in active development, notably CARTA\footnote{https://cartavis.github.io/}
\citep{carta2018} and
SlicerAstro \citep{Punzo2017}.

\IDAVIE\ is still very limited in functionality compared to some of these much matured
tools, and from these we are aware of highly useful functionality that would be
desirable to our \IDAVIE\ users.  For example, iso-contours are extremely powerful graphic
constructs that can reveal hidden features in particle-dense or volumetric-bright regions of a
data set or cube.  We will continue to develop \IDAVIE\ to incorporate these methods and tools,
as well as innovate how they are deployed given the unique requirements of 3D immersion.

\paragraph{Versatility of \IDAVIE:}Not only astronomy focused, we have worked with cellular and molecular biologists, 
neuroscientists, eResearch data scientists, and even graphic and fine artists who want to use the immersive environment.  Hence, our software development has general application, and we strive to be flexible to 3D data sets coming from all walks of research and applied sciences.   Thus far, most of our development has been toward volumetric rendering, and notably spectral cubes -- but any volume can be explored with \IDAVIE.  

\paragraph{Public Release of \IDAVIE--v:}We have released a ``beta" version of this system to the radio astronomy community
\citep{angus_comrie_2021_4614116},
but we hasten to add that the system works generally for any volumetric data as long as the input is in FITS format.  Development has not ceased with this release;  we continue to improve and add upon these tools and functionality, but we are also working on cutting-edge methods for doing better \VR.  We describe some of this current and future work below, including an innovative combination of \VR\ in the full-dome planetarium,  and \VR\ through cloud streaming.  We start with, however, the challenges with \VR\ that we have overcome or currently grappling with.

\subsection{Current and ongoing Challenges: }

There are many challenges with \VR\ technology, such as gaming engines and the controller software that act as 
barriers to broad adoption. Some are more serious (read: chronic) than others,  which we believe with time, technology and software enhancements will cease to be barriers.  First up is the computer requirements and the headsets.  

High-end GPUs are a must for research using \VR, the reason is that the data is large (millions of particles, GB-class cubes), visual fidelity is premium and the refresh rate is taxing (typically 90hz).  In the IVL, we have used NVIDIA 1, 2 and 3-series, 
all of which are pricey.  To date, only Windows-based drivers are viable, the speed and power that is required for high-end research. Astronomers do not generally use the Windows-OS, rather preferring open-source Linux or even the Mac-OS.  Hence, purchasing, configuring and maintaining Windows machines is not a straight-forward experience for most astronomers (but perhaps less of a barrier for other disciplines and applied fields).  Linux drivers are getting better and we foresee the day in the near future when our software tools will migrate over to Linux, or virtual Linux machines embedded within Windows.

For the headset, the premium is on pixel resolution for each eye, currently about 2K, with higher resolutions in the pipelines of manufacturers. High pixel fidelity and large ($>$110 deg) field-of-view is most welcome, but also puts pressure on the GPU performance (notably an issue with gaming laptops, where weight, temperature and size constraints limit the processing power).  We have even found that the hand controller design makes a big difference with both comfort and efficiency.  Our system deploys as many voice commands as is practical (several dozen) in order to mitigate the limitations of hand controllers and \VR\ menu interaction (see Appendix C).  There is also the chronic problem of the \VR\ view being slightly out of focus, as the headset optics are pushed hard to provide field-of-view and high resolution with the GPU power and bandwidth available.  Fidelity is a must with research, discerning fine details in data is often the path for discovery.

The next challenge is the framerate required for user comfort. Generally the minimum is 75 to 90\,Hz, while higher rates are more desirable.   These rates are not always maintained when data volume approaches or exceeds what the hardware is capable of doing.  Dropped frames, resulting in a halting motion, is very uncomfortable for the user.  Hence, the more power (GPUs), bandwidth and core memory, the better viewing, as rendering is heavily dependent on the data volume, for example.   

Limitations due to core (CPU) memory have been a challenge for our large data sets.  As described in \S\,4, we have overcome much of this limitation by down-sampling $>$Gb cubes, and then up-sampling to native resolution when a smaller region of the cube is selected.  GPU memory is not as big a problem because of the software, Unity gaming engine e.g., limitation of only being able to render relatively small cubes (less than a GB) at one time.  This may not be a limitation for other rendering engines, such as those from open-source -- and hence a good option to explore in the near future.   GPUs are getting faster and more powerful, but they are also improving with their own memory bandwidth, which ultimately limits the speed of data transfer from CPU to GPU 
\citep[see e.g.,][]{Biedert2018}.

A variety of smaller challenges are associated with \VR\ research.    Setting up a separate space where the computer and sensors are located, and where the user is able to stand and move around in a limited fashion (tether and obstacle bound).  The tether or wire limitation may someday be fully replaced by wireless streaming ; see below where we discuss the viability of streaming \VR.
Of course,  some headsets have built-in sensors making for a compact user environment -- the user can just as easily sit in a comfortable chair and use \VR\ to its full capabilities with good hand control motions.  The headsets have greatly improved, but it is still not as smooth to take on and off the headset, moving from the \VR\ environment to the computer screen/keyboard environment and back (we find that our computer interfaces are just as important as our \VR\ interfaces, both are often needed).  The gaming engines, such as Unity, have limited and relatively un-friendly interfaces, not well suited for research; hence, we have designed our own, User Interface, to be more intuitive and useful to the researcher.  An example of the interface and menus used for the volume rendering mode is provided in the Appendix C.  

Another user challenge, the lack of a reliable and high-fidelity mouse and keyboard in the \VR\ environment.
The solution will be hand controllers that recognize high-fidelity gestures, such has finger motions (e.g., typing on a virtual keyboard).  We have had limited success with these early prototypes, and await the next generation of controllers and trackers.   
Technology and software development are improving and at a fast rate, slowly removing or mitigating these challenges and limitations. 

Finally, as we have noted early on, the choice of colour for data transformation and visualisation is quit important to optimally distinguish, contrast and discern features.   
The
colour tables available in matplotlib are not always optimal, there are better transforms being actively developed and created by other groups.  
Of note is that there are colour schemes that both create spatial depth in 2-D and avoid being as misleading as a rainbow colour map is for some use cases  \citep[see][]{English2017CanvasAC}).  Our lab will continue to explore colour and visual perception through the 3D \IDAVIE\ system.

\subsection{The Way Forward}

The \IDAVIE\ system is a great leap forward with research using \VR\ technology, yet it is still relatively primitive compared to matured and time-tested tools and applications (e.g., compare with SlicerAstro, a powerful 3D rendering tool set).  We will continue to add functionality, such as 
iso-contours,  or 
the ability to overlay other cubes (of similar size, or cubelets or sub-cubes), which may be a completely different wavelength and data set; e.g., comparing two different emission lines, such as \HI\ and molecular CO.  

We strive for better user feedback with interactive menus and graphics windows within the \VR\ environment, including more efficient hand control gesturing, audio and haptic feedback. Other useful functions include  access to clouds and archival data within the \VR\ environment, pulling physical and 2D information in real-time that helps understand what you are rendering in 3D (e.g., the M31 example shown in Fig~\ref{fig:m31snap}).  There are many other features, inspired by KVIS and SAOIMAGE-ds9 for example, that we want to incorporate into our system.   From the group interaction and collaboration realm,
a popular request, and clearly a powerful one, is to have multi-user capability (or what is referred to as multi-player in the gaming field).  Having the ability to converse and interact with another user in the same \VR\ environment, whether they are next to each other, or located continents apart,  opens up rich collaboration and more efficient ways of exploring data. \\

\textbf{Streaming with \VR:} Perhaps the most exciting future we can look forward to with \VR\ is the ability to stream our data and calculations from a cloud or server (which may be in the same room or offsite) and thus decouple our reliance on our own computing hardware and storage demands.   Current technology has this capability (e.g., using the Oculus Quest 2), but in limited form that is not yet practical to scientific exploration of large data sets.   However, our initial experiments with wire-less cloud-based streaming have highlighted the way forward.  Here we describe our vision of streaming VR.

\begin{figure}[h!]
\vspace{-5pt}
\includegraphics[width=0.50\textwidth]{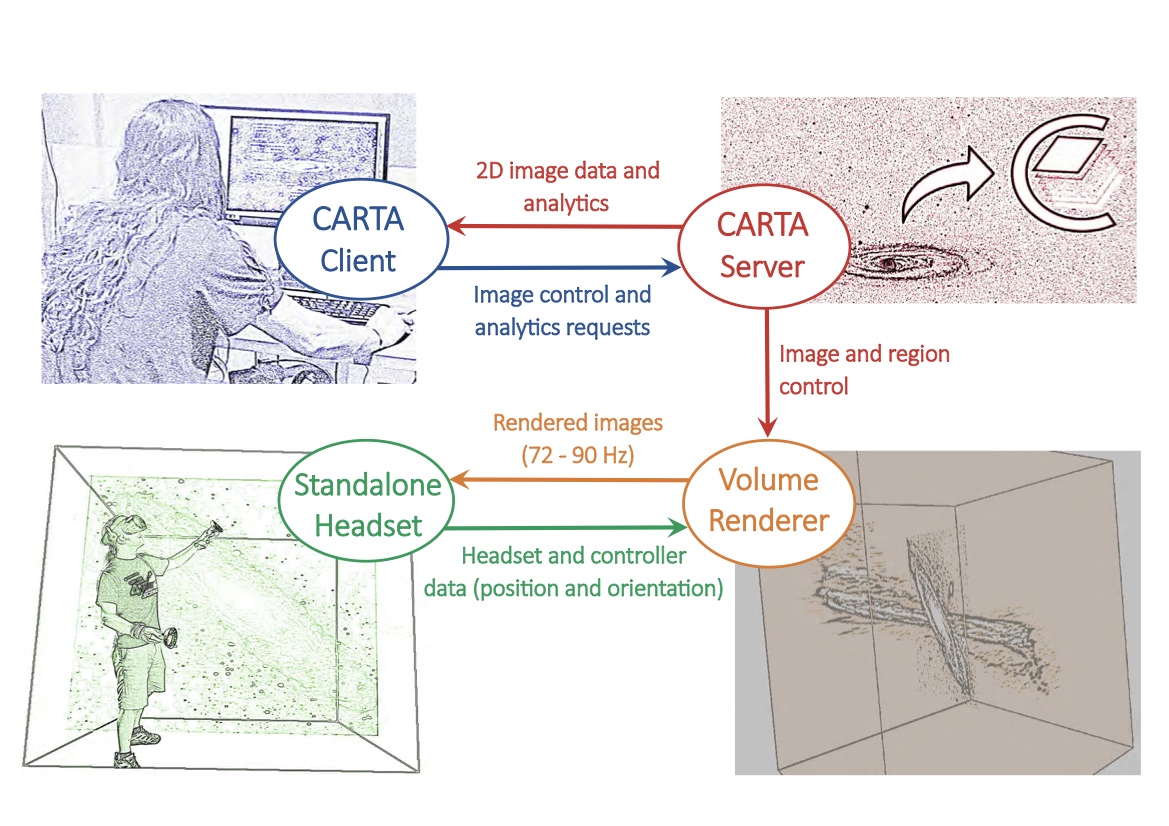}
\centering
\vspace{-25pt}
\caption{\footnotesize{
Basic concept of stream-VR, starting from a regular 2D terminal using 3D rendering software (such as CARTA streaming from a cloud server), transforming to \VR\ ready viewing.  
}}\label{fig:stream}
\vspace{-5pt}
\end{figure}

It would be cloud-based rendering, using a high performance computer with a fast and high bandwidth connection to the internet. There would be access to far more GPU power and CPU memory volume this way, all of which would use efficient multi-thread coding. 
It would be scalable to much larger datasets, with $>$50Gb cubes no longer a challenge to render.   Headset hardware would be more diverse, only requiring a hand tracking sensor(s) and a modest CPU/GPU in the headset, and hence would be much simpler in design and cost (e.g., the Oculus Quest 2 would work well with this system).  Fast internet connection to the internet would be required for the user to minimize deleterious latency.   A major bonus:  there would be less reliance (and far more flexibility) with operating systems.  

Control and menu interaction could be with a simple browser, and thus much less reliance on custom software, and so the interface design is simplified. Adding new analysis tools, through the web interface, would be straight forward, and in principle, inter-operability with existing tools (e.g., CARTA) would be possible; see our concept illustrated in Fig.~\ref{fig:stream}. 
Of course, there are some major challenges to streaming \VR, not the least of which is latency and internet bottlenecks, which may render very uncomfortable viewing or limiting what can be viewed. Premiums will be placed on GPU-accelerated encoding and decoding 
\citep{Biedert2018},
latency compensation (e.g., using artificial intelligence to smoothly fill latency or dropped-frame gaps) and
other clever designs that are not anticipated at this time.

There is great promise with cloud-based visualisation, and our lab will be actively working on this in the coming years.  To date, we have already made progress with streaming small data cubes from a local ``server" to a standalone wire-less \VR\ headset (in this case, the Oculus Quest 2).  We are specifically experimenting with better down-sampling and compression methods (from the server side) and more efficient rendering techniques from the client side.


\subsection{\VR\ to Dome, \IDAVIE--d} 

We finish with a novel project that combines \VR\
research and the full-dome (planetarium) environment.
This is not to be confused with \VR\ headsets used by
the audience in a dome or otherwise;  rather it is
the researcher (or presenter) who dons the headset
in front of the audience 
and interacts with spectral imaging
data using \IDAVIE--d.  The presenter's view is seen
from a second virtual camera (in the \VR\ environment) and
projected in real-time onto the dome surface with a proper fisheye distortion to match the curved dome.
In this way, the theatre audience is able to see 
and follow the rendering that is happening in the \VR, and through 
full 360-degree immersion that domes enable to large audiences.  
This overcomes the major limitation of \VR\ exploration being
experienced by only the user (or perhaps those standing around
the flat display computer). Here you are only limited by the size
of the theatre. Planetarium presenters also benefit from the more intuitive and natural motion controls of \VR\ over the traditional keyboard-mouse or joystick interface that is used to ``fly" through datasets in planetarium fashion.

There are two ways to link the \VR\ system to the (planetarium) software that controls the digital projection.  The first is simply using the video capture system that the projection software may have (assuming the hardware is installed, and the corresponding software drivers are configured for the dome specifications).  The output from a virtual spectator camera in the \VR\ scene is then piped through the video capture and projected onto the dome.  It may be limited by the resolution of the video capture cards (e.g., 2K resolution), which can result in a fuzzy view to the audience.   The second option is to load \IDAVIE\ directly onto the computers of the projection system and exploit the full resolution that can be properly multi-projected and blended to the dome.  The difficulty here is enabling our software use of the planetarium's cluster rendering architecture.  In our experimentation, we have only tried the first method -- using the Iziko Planetarium facility (with 2K video capture), and the Colgate University Ho Tung Visualization Lab (with 4K video capture) as part of the Data2Dome Workshop in October of 2019. 

\begin{figure}[h!]
\includegraphics[width=0.48\textwidth]{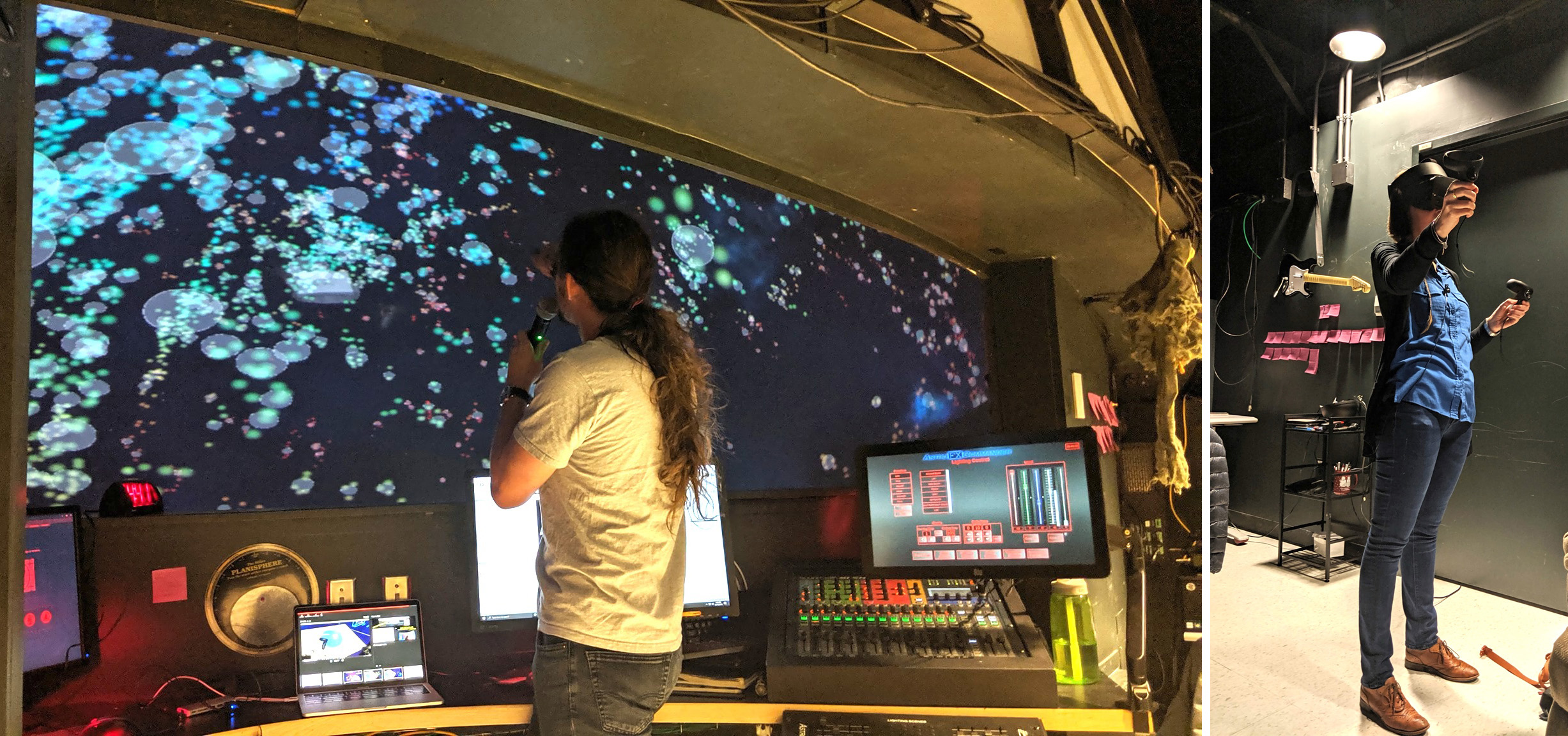}
\centering
\vspace{-20pt}
\caption{\footnotesize{
The humans behind \VR\ to Dome, in action at Colgate University in October of 2019.  On the (right) with the \VR\ headset, Lucia Marchetti, presenting the data to the theatre audience (note the headset graphics cable leading to the computer room); (left)  is PhD student, Alex Sivitilli, at the control panel coordinating the \VR\ to Dome projection and oral presentation.
}}\label{fig:VR2}
\vspace{-5pt}
\end{figure}
Fig.~\ref{fig:VR2} shows a live demonstration of \IDAVIE--d in an audience-filled theatre, where the \VR\ presenter with headset and hand controllers is standing close to the computer room so that a direct cable connection is made with the projection computer cluster, and adjacent (left) is the actual projection control computer which is coordinating the presentation.  The physical connection need not be a short wire restricting motion of the presenter, but could be a wire/cable that extends to the front of the theatre where the presenter could stand and describe what the audience is seeing, or even a wireless system through a stream-VR-capable system.

\begin{figure}
\includegraphics[width=0.48\textwidth]{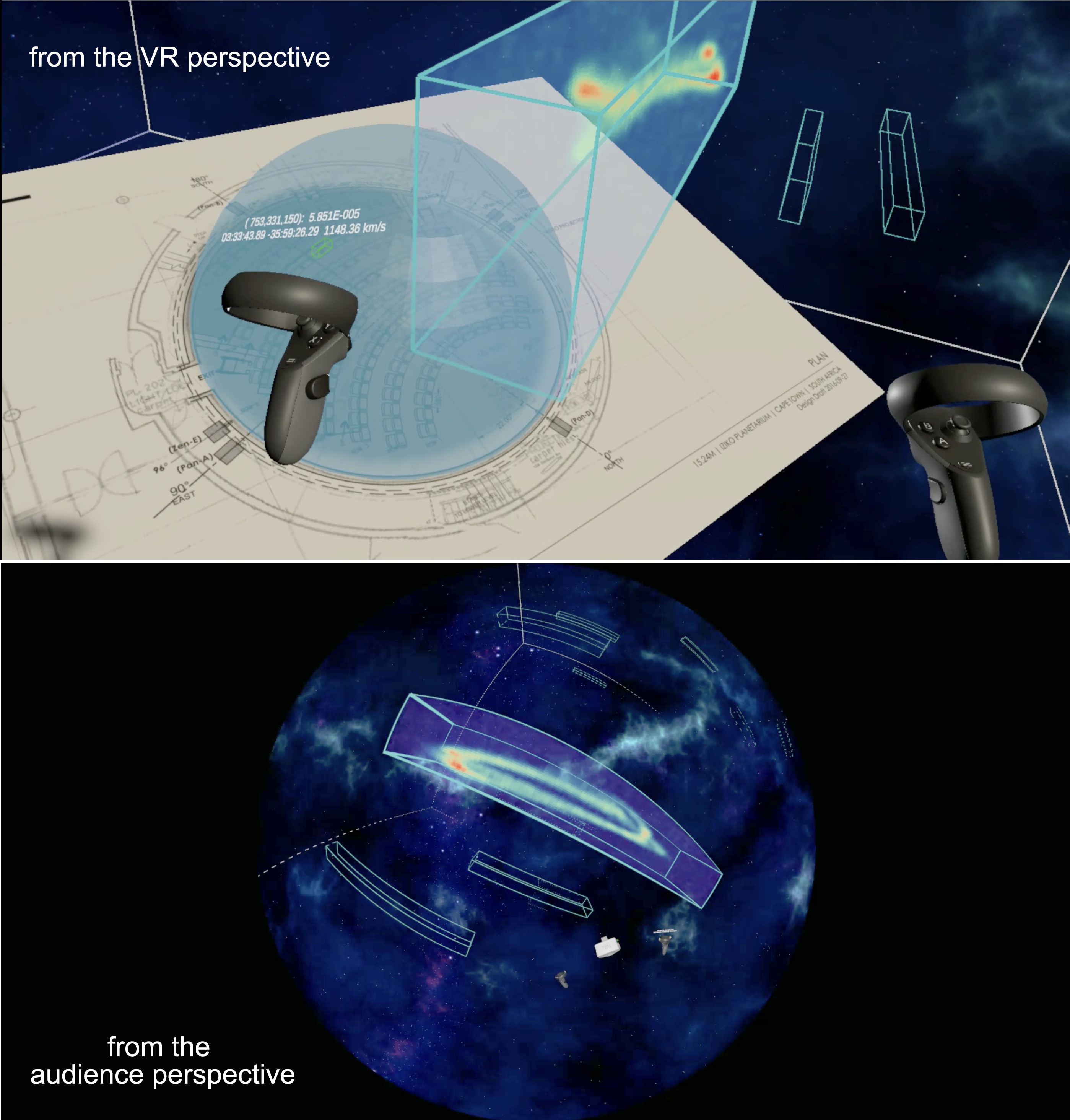}
\centering
\vspace{-20pt}
\caption{\footnotesize{
\VR\ to Dome in the Colgate University planetarium, where the \VR\ presenter is demonstrating to an audience how \IDAVIE\ renders and interacts with an \HI\ spectral cube; here extragalatic sources (most notably galaxy NGC\,1365) in the Fornax Galaxy Cluster are marked with blue boxes.
The top panel is the \VR\ presenter's view, showing the 
3D rendering through the headset (from one eye), and is moving the data around with respect to the ``virtual dome" (top panel).  The audience (bottom panel) is looking up at the dome projection and views the rendering in real time.  
}}\label{fig:VR3}
\vspace{-10pt}
\end{figure}

During this workshop and \IDAVIE--d demonstration, we showcased two of our data sets:  the 2MRS galaxy and groups catalogue (large scale structure in the local universe; see \S\,4.1) through \IDAVIE--p, and the Fornax Galaxy Cluster \HI\ spectral volumes through \IDAVIE--v.  Fig.~\ref{fig:VR3} shows still images from one of these presentations.  The top panel is the presenter's eye within the \VR\ environment, working with the Fornax spectral cube, moving it and sizing it over a virtual ``dome", representing the theatre and audience within. This is essentially a ``gods eye" view of the dual presentation, the presenter looking down at the dome and in full control of how the audience see the data.  

Meanwhile, the bottom panel (\ref{fig:VR3}) is the view from the audience, as they relax in their reclining seats looking up at the dome and digital projection.  They can see the facemask and hand controllers of the presenter in the foreground of the data, while the volumetric data appears with the \IDAVIE--v interaction tools (the \HI\ sources highlighted with blue boxes, featuring the NGC\,1365 galaxy).  The presenter is able to work all of the functionality of \IDAVIE, and the audience is able to see it all in a 360-degree immersive environment.  VR2Dome is just one of the innovative and transformative developments that the IVL is conducting in response to the Big Data and advanced visualisation era we are now fully engaged with.

\vspace{10pt}
 
  \textbf{Final Thoughts: }
 Over the past several years, as we have worked with our visualisation tools and research colleagues, we have seen rapid evolution -- both in in visualisation tools and techniques, but also in the arrival of Big Science, with its correspondingly massive and complex datasets. We feel the investment in the IVL has been both fruitful and opportune -- its origin and (growing) success owing to a unique and synergistic collaboration between academic researchers and research institutions (e.g., University of Cape Town; University of Groningen, INAF), data intensive centers (e.g., IDIA), and government funding and support (e.g., National Research Foundation of South Africa). This paper represents the first detailed report on our development in the IVL, providing reports on our \IDAVIE\ system that uses \VR\ technology in conjunction with commercial software (such as the gaming engine, Unity), and our own suite of custom software tools that enable user-interaction with 3D data sets, exploiting the boundless human capacity to visually comprehend and perceive complex phenomena. We have shown how \VR\ R \& D is benefitting research on galaxy evolution, cosmic web large-scale structure, galaxy-galaxy interactions, and gas/kinematics of nearby galaxies in survey and targeted observations. It reveals the vast potential, and largely still hidden capabilities of this new technology, and we will continue to develop and push the frontier forward.





\section*{Acknowledgements}

Special thanks to Trystan Lambert, Nathan Deg, Meridith Joyce, Claude Carignan,  Gyula Jozsa, Marcin Glowacki,  Julia Healy, and Charl Cater for assisting with testing and validation of the \IDAVIE\ system.  We thank St\'{e}phane Courteau, Mike Hudson, Mark Subbarao, Jayanne English, Miguel Aragon-Calvo,  Mark Neyrinck and Chris Fluke for insightful discussion on the future of VR and visualisation in data science.
We thank our interdisciplinary colleagues Ben Loos and Andre Du Toit for fascinating visualisations from the very small (molecular biology) angular-scale world.
Special thanks to Joseph Eakin for wonderful technical service during our VR-to-Dome demonstration at Colgate University.  
And finally a huge thanks to the two referees who provided a number of excellent suggestions and critiques that helped sharpen the manuscript.
 THJ acknowledges funding from the National Research Foundation under the Research Career Advancement and South African
 Research Chair Initiative programs (SARChI), respectively.
This work made use of the Inter-University Institute for Data Intensive Astronomy (IDIA) Visualization Lab\footnote{ https://vislab.idia.ac.za}. IDIA is a partnership of the University of Cape Town, the University of Pretoria and the University of the Western Cape. The authors acknowledge financial support from the Italian Ministry of Foreign Affairs and International Cooperation (MAECI Grant Number ZA18GR02) and the South African
NRF (Grant Number 113121) as part of the ISARP RADIOSKY2020 Joint Research Scheme.  
This project has received funding from the European Research Council (ERC) under the European Union’s Horizon 2020 research and innovation programme (grant agreement no. 679627; project name FORNAX).






%






\clearpage


\appendix

\twocolumn

\section{\IDAVIE\ Mapping Files}\label{appendixa}

JSON mapping schema for loading tables and catalogues.

\vspace{+10pt}

\begin{lstlisting}[language=json]
{
  "$schema": "http://json-schema.org/draft-06/schema#",
  "definitions": {
    "ColorMap": {
      "enum": [
        "Jet",
        "Magma",
        "Plasma",
        "Prism",
        "Rainbow",
        "Viridis",
      ],
      "type": "string"
    },
    "RenderType": {
      "enum": [
        "Billboard",
        "Line"
      ],
      "type": "string"
    },
    "ShapeType": {
      "enum": [
        "Halo",
        "Circle",
        "OutlinedCircle",
        "Square",
        "OutlinedSquare",
        "Triangle",
        "OutlinedTriangle",
        "Star"
      ],
      "type": "string"
    },
    "ScalingType": {
      "enum": [
        "Linear",
        "Log",
        "Sqrt",
        "Squared",
        "Exp"
      ],
      "type": "string"
    },
    "MapFloatEntry": {
      "properties": {
        "Clamped": {
          "type": "boolean"
        },
        "MinVal": {
          "type": "number"
        },
        "MaxVal": {
          "type": "number"
        },
        "Offset": {
          "type": "number"
        },
        "Scale": {
          "type": "number"
        },
        "ScalingType": {
          "$ref": "#/definitions/ScalingType"
        },
        "Source": {
          "type": "string"
        }
      },
      "required": [
        "Source"
      ],
      "type": "object"
    },
    "MappingUniforms": {
      "type": "object",
      "properties": {
        "ColorString": {
          "type": "string",
          "pattern": "^#([a-fA-F0-9]{6}|[a-fA-F0-9]{3})$",
          "example": "#aa33cc"
        },
        "Scale": {
          "type": "number"
        },
        "PointSize": {
          "type": "number"
        },
        "Opacity": {
          "type": "number"
        },
        "PointShape": {
          "$ref": "#/definitions/ShapeType"
        }
      },
      "additionalProperties": false
    }
  },
  "properties": {
    "ColorMap": {
      "$ref": "#/definitions/ColorMap"
    },
    "Spherical": {
      "type": "boolean"
    },
    "RenderType": {
      "$ref": "#/definitions/RenderType"
    },
    "UniformColor": {
      "type": "boolean"
    },
    "UniformPointSize": {
      "type": "boolean"
    },
    "UniformOpacity": {
      "type": "boolean"
    },
    "UniformPointShape": {
      "type": "boolean"
    },
    "Uniforms": {
      "$ref": "#/definitions/MappingUniforms"
    },
    "Mapping": {
      "properties": {
        "X": {
          "$ref": "#/definitions/MapFloatEntry"
        },
        "Y": {
          "$ref": "#/definitions/MapFloatEntry"
        },
        "Z": {
          "$ref": "#/definitions/MapFloatEntry"
        },
        "X2": {
          "$ref": "#/definitions/MapFloatEntry"
        },
        "Y2": {
          "$ref": "#/definitions/MapFloatEntry"
        },
        "Z2": {
          "$ref": "#/definitions/MapFloatEntry"
        },
        "Lat": {
          "$ref": "#/definitions/MapFloatEntry"
        },
        "Lng": {
          "$ref": "#/definitions/MapFloatEntry"
        },
        "R": {
          "$ref": "#/definitions/MapFloatEntry"
        },
        "Cmap": {
          "$ref": "#/definitions/MapFloatEntry"
        },
        "Opacity": {
          "$ref": "#/definitions/MapFloatEntry"
        },
        "PointSize": {
          "$ref": "#/definitions/MapFloatEntry"
        },
        "PointShape": {
          "$ref": "#/definitions/MapFloatEntry"
        }
      },
      "additionalProperties": false,
      "type": "object"
    },
    "metaMapping": {
      "properties": {
        "Name": {
          "properties": {
            "Source": {
              "type": "string"
            }
          },
          "additionalProperties": false,
          "required": [
            "Source"
          ],
          "type": "object"
        }
      },
      "additionalProperties": false,
      "type": "object"
    }
  },
  "additionalProperties": false,
  "required": [
    "Mapping"
  ],
  "type": "object"
}
\end{lstlisting}

\newpage

\onecolumn

Example of a mapping JSON file for loading a redshift catalogue.

\vspace{+10pt}

\begin{lstlisting}[language=json]
{
  "RenderType": "Billboard",
  "ColorMap": "Binary",
  "Spherical": true,
  "UniformOpacity": true,
  "UniformPointShape": true,
  "UniformPointSize": false,
  "Uniforms": {
    "ColorString": "#FFFFFF",
    "PointShape": "Halo",
    "Scale": 0.001
  },
  "Mapping": {
    "Lat": {
      "Source": "glon"
    },
    "Lng": {
      "Source": "glat"
    },
    "R": {
      "Source": "Dm"
    },
    "Cmap": {
      "Source": "Dm",
      "Scale": 0.0015
    },
    "PointSize": {
      "Source": "zphoto",
      "Scale": 10
    }
  }
}
\end{lstlisting}

\newpage

\onecolumn

\section{\IDAVIE--v Sequence Diagram}\label{appendixb}


\begin{figure*}[h!]
\centering
\includegraphics[width=0.46\textwidth]{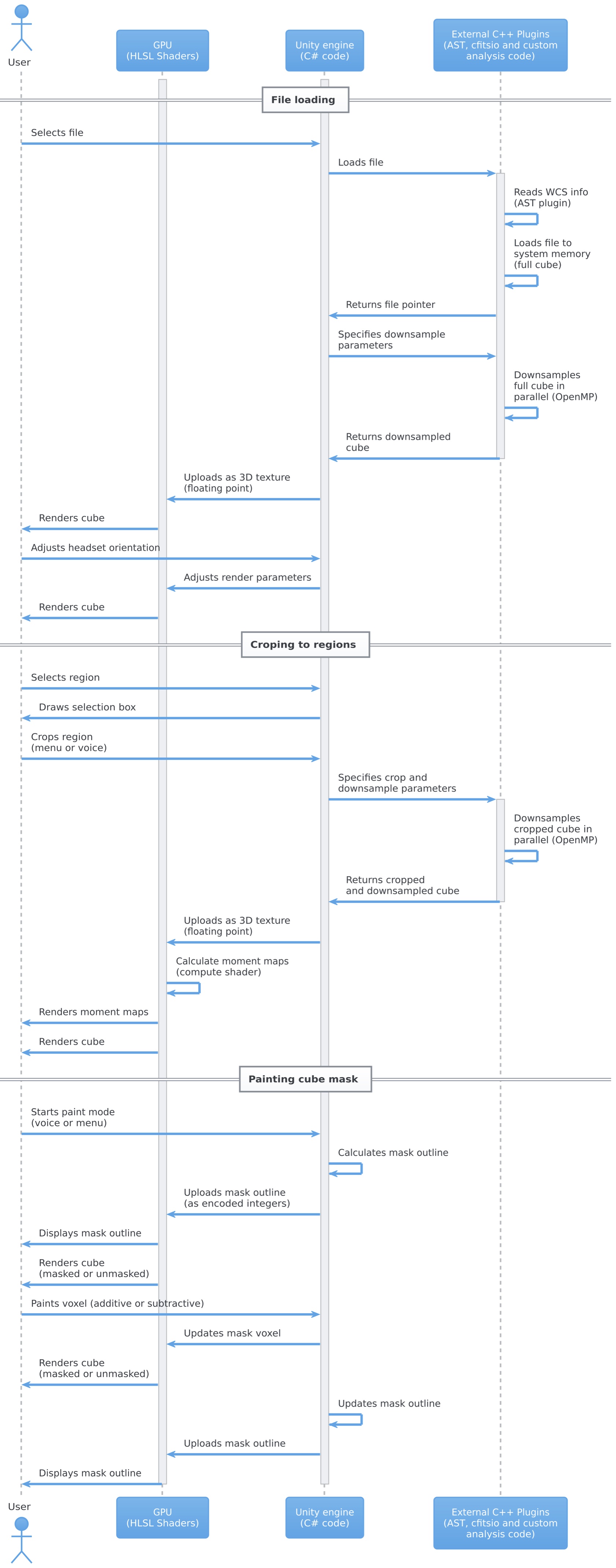}
\centering
\vspace{-5pt}
\caption{\footnotesize{\IDAVIE--v System   
sequence diagram for
 (top) loading a file, (middle) cropping to a selected region and (bottom) painting voxels in the mask cube. See \S\,3.
}}\label{fig:sequence}
\vspace{-0pt}
\end{figure*}

\newpage

\onecolumn

\section{\IDAVIE\ Graphics Menus}\label{appendixc}

\IDAVIE--v has interactive graphics windows as part of the start-up data load procedure, and within the \VR\ environment.  Examples of the graphics menus are given in this appendix.

\begin{figure*}[h]
\includegraphics[width=1\textwidth]{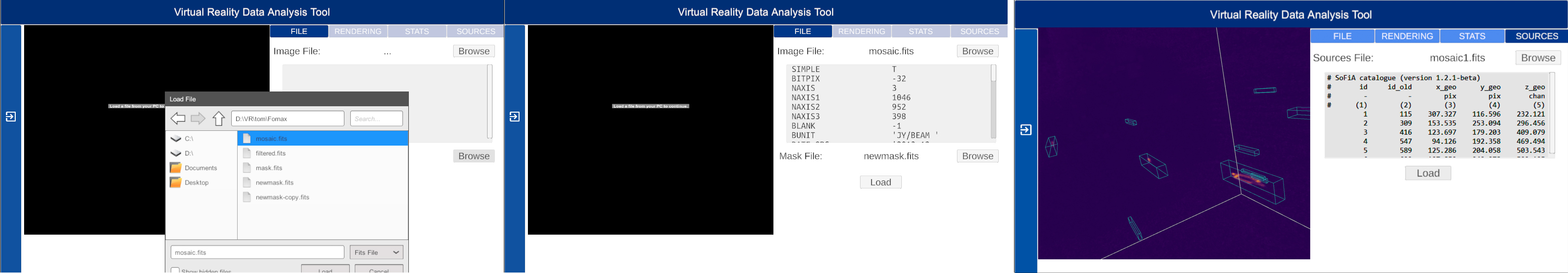}
\centering
\vspace{-10pt}
\caption{\footnotesize{
User interface for \IDAVIE--v.  Any compliant FITS cube may be loaded with ``Image File".  Optionally, a ``mask" file (cube of the same dimensions as the primary image) may also be loaded.   The FITS header is displayed (center panel) after the cube is loaded.   The user has the option of loading a source table; example shown right panel of a SoFiA table.  Sources ``boxes" demarcate the location of sources in the table as rendered in \IDAVIE.
}}\label{fig:browser}
\vspace{-10pt}
\end{figure*}

\begin{figure*}[h!]
\vspace{-5pt}
\includegraphics[width=.85\textwidth]{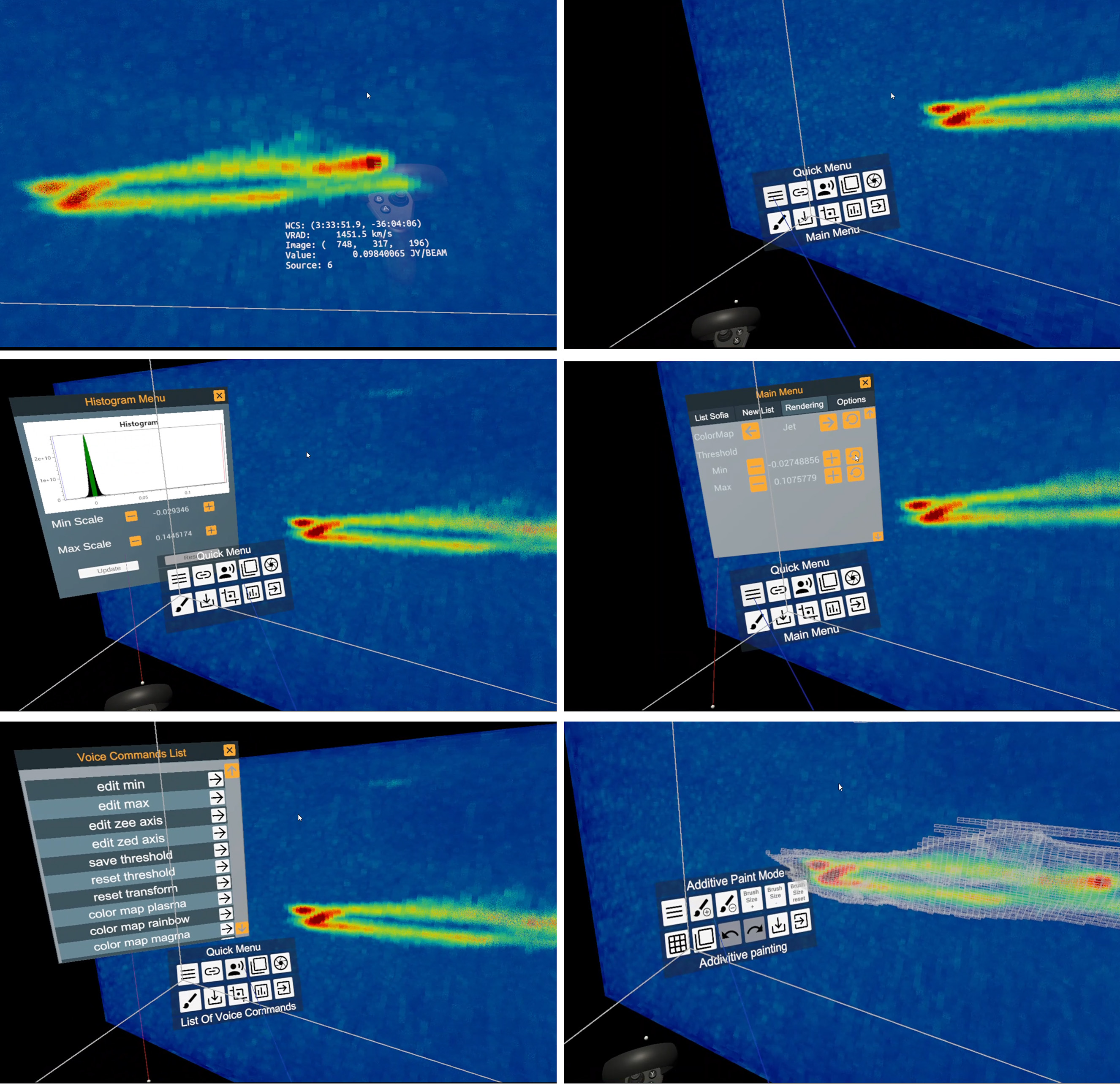}
\centering
\vspace{-10pt}
\caption{\footnotesize{
Inside the \VR\ environment, the \IDAVIE--v coordinate feedback and interactive menus.  The Quick Menu (upper right) in invoked with the hand controller (holding down Button A), which has a number of useful functions (e.g., statistics, toggle masks, cropping, snap a picture, list of voice commands, paint mode).
The Main Menu (with additional functions) may also be selected from the Quick Menu.
}}\label{fig:menus}
\vspace{-10pt}
\end{figure*}


\twocolumn

\typeout{get arXiv to do 4 passes: Label(s) may have changed. Rerun}

\end{document}